\documentclass[twocolumn,showpacs,preprintnumbers,amsmath,amssymb]{revtex4-1}
\usepackage{graphicx}
\usepackage{dcolumn}
\usepackage{bm}
\usepackage{braket}
\usepackage{ulem}
\usepackage[usenames,dvipsnames]{color}% use as \textcolor{red}{******} 
\usepackage[%dvipdfmx,
colorlinks=true, citecolor=blue, urlcolor=blue, linkcolor=blue,
setpagesize=false, bookmarks=false
]{hyperref}

\usepackage[english]{babel}
\addto{\captionsenglish}{}

\begin{document}

\title{
Photoinduced $\mbox{\boldmath{$\eta$}}$ pairing in the Kondo lattice model
}

\author{
  Tomonori Shirakawa$^{1}$,
  Shohei Miyakoshi$^{2}$, and
  Seiji Yunoki$^{1,2,3}$
} 
\affiliation{
${}^1$Computational Materials Science Research Team, RIKEN Center for Computational Science (R-CCS), Kobe, Hyogo 650-0047, Japan \\
${}^2$Computational Quantum Matter Research Team, RIKEN Center for Emergent Matter Science (CEMS), Wako, Saitama 351-0198, Japan \\
${}^3$Computational Condensed Matter Physics Laboratory, RIKEN Cluster for Pioneering Research (CPR), Wako, Saitama 351-0198, Japan
}

\date{\today}

\begin{abstract}
  The previous theoretical study has shown that pulse irradiation to the Mott insulating state in the Hubbard model 
  can induce the enhancement of superconducting correlation due to the generation of $\eta$ pairs 
  [Kaneko {\it et. al.,} Phys. Rev. Lett. {\bf 122}, 077002 (2019)]. 
  Here, we show that the same mechanism can be applied to the Kondo lattice model, an effective model
  for heavy electron systems, by demonstrating that the pulse irradiation indeed enhances the $\eta$-pairing correlation. 
  As in the case of the Hubbard model, the non-linear optical process is essential to increase the
  number of photoinduced $\eta$ pairs and thus the enhancement of the superconducting correlation. 
  We also find the diffusive behavior of the spin dynamics
  after the pulse irradiation, 
  suggesting that the increase of the number of $\eta$ pairs leads to the decoupling between the conduction band 
  and the localized spins in the Kondo lattice model, 
  which is inseparably related to the photodoping effect. 
\end{abstract}

\maketitle

%{\bf\color{blue} ========================== }

\section{Introduction}

Recent extensive studies of photoinduced states of strongly correlated materials have paved the way
to find new states of matter~\cite{tokura2006,iwai2006,giannetti2016}.
Indeed, they have observed many intriguing phenomena, including
the photoinduced transient superconducting behavior~\cite{Fausti189,Hu:2014aa,PhysRevB.89.184516,Mitrano:2016aa,Cantaluppi:2018aa}
and the photoinduced insulator-to-metal transition~\cite{PhysRevLett.91.057401,PhysRevLett.98.037401,doi:10.1143/JPSJ.77.113714,PhysRevB.82.060513,PhysRevB.83.125102}. 
These experimental observations have stimulated theoretical studies on non-equilibrium dynamics of strongly correlated electrons, 
mostly focusing on photoexcited states in Hubbard-like models~\cite{YONEMITSU20081,RevModPhys.86.779,ishihara2019,oka2019}.

In this context, we have previously studied a photoexcited state after pulse irradiation onto 
the Mott insulating state in the Hubbard model and found the strong enhancement of superconducting correlation
due to $\eta$ pairing~\cite{PhysRevLett.122.077002}, which is a pair density wave with phase $\pi$ and is associated with 
the transverse components of pseudo-spin $1/2$ operators, first introduced 
by C.~N.~Yang~\cite{PhysRevLett.63.2144}.
We have also shown that the $\eta$ pairs are preferentially excited by the optical pulse field 
because of the selection rule forced by the symmetry of the $\eta$-pairing 
operators~\cite{PhysRevLett.122.077002,fujiuchi2019,kaneko2020}.

In this paper, we report that the same mechanism 
can be applied to another class of models, the Kondo lattice model,  
known as an effective model to describe
electronic states in heavy electron systems~\cite{tsunetsugu1997,cox1998}. 
The Kondo lattice model is composed of a conduction band with mobile electrons and 
localized spins coupled to each conduction site antiferromagnetically. 
Because the presence of the exchange interaction induces
a nontrivial scattering between the mobile electrons and localized spins, 
the Kondo lattice model is studied in the context of many-body quantum systems~\cite{tsunetsugu1997}.

We demonstrate numerically that the pulse irradiation onto 
the Kondo insulating ground state induces the enhancement
of the $\eta$-pairing correlation. The enhancement of the $\eta$-pairing 
correlation is due to the increase of the number of  
$\eta$ pairs that are selectively generated by the pulse optical field 
because the Kondo lattice model possesses the pseudo-spin $1/2$, i.e., $\eta$-SU(2), 
symmetry under which the current operator is a rank 1 tensor operator. 
This implies that a non-linear optical process is essential to 
increase the number of $\eta$ pairs and thus the enhancement of the superconducting correlation. 
Moreover, we find that the spin dynamics changes drastically 
and becomes diffusive after the pulse irradiation. This can be understood 
because the generation of $\eta$ pairs by the pulse irradiation is 
equivalent to in-situ doping of carriers for spin dynamics.

The rest of this paper is organized as follows. 
We first introduce the Kondo lattice model on a bipartite lattice
and the $\eta$-pairing operators in Sec.~\ref{sec:method}. 
We also explain how to introduce the pulse optical field into the model 
and discuss the effect 
with the time-dependent perturbation theory. 
We then show our numerical results in Sec.~\ref{sec:result} 
and conclude this paper with a brief discussion 
in Sec.~\ref{sec:summary}. The numerical details are supplemented in Appendix~\ref{appendix:spec}

\section{$\mbox{\boldmath{$\eta$}}$ pairing in Kondo lattice} \label{sec:method}

In this section, we describe the $\eta$-pairing and the photoexcitation in the Kondo lattice model.
We first introduce the Kondo lattice model on a bipartite lattice in Sec.~\ref{sec:model}.
We next introduce the pseudo-spin 1/2 operators, i.e., 
$\eta$-pairing operators, defined for the Kondo lattice model in Sec.~\ref{sec:etaop}. 
We then show that the Kondo lattice model possesses the $\eta$-SU(2) symmetry and 
discuss its consequences.  
In Sec.~\ref{sec:tensor}, we introduce two kinds of tensor operators relevant to the photoexcitation process.
We describe how to introduce the time-dependent field into the model in Sec.~\ref{sec:simulation}
and analyze the effect with the time-dependent
perturbation theory in Sec.~\ref{sec:perturb}.

\subsection{\label{sec:model}Model}

The Kondo lattice model is described by the following Hamiltonian: 
\begin{equation}
   \hat{\mathcal{H}} = \hat{\mathcal{H}}_t + \hat{\mathcal{H}}_J, \label{eq:ham},  
\end{equation}
where 
\begin{equation}
   \hat{\mathcal{H}}_t = - t \sum_{\langle j, j^{\prime} \rangle} \sum_{\sigma = \uparrow,\downarrow} 
  ( \hat{c}_{j \sigma}^{\dagger} \hat{c}_{j^{\prime}\sigma} + \hat{c}_{j^{\prime}\sigma}^{\dagger} \hat{c}_{j\sigma} )
  \label{eq:ham:hop} 
\end{equation}
and 
\begin{equation}
 \hat{\mathcal{H}}_J = J \sum_{j} \hat{{\bm S}_j} \cdot \hat{{\bm M}_j}.
  \label{eq:ham:ex}
\end{equation}
Here $\hat{c}_{j\sigma}$ ($\hat{c}_{j\sigma}^{\dagger}$) denotes the annihilation (creation) operator
of a mobile electron with spin $\sigma$ (=$\uparrow,\downarrow$) at site $j$. 
The first sum in Eq.~(\ref{eq:ham:hop}) indicated by $\langle j, j^{\prime} \rangle$ runs over  
all pairs of nearest-neighbor sites $j$ and $j^{\prime}$ in the lattice. 
We assume that the lattice is bipartite in which sites 
can be divided into two sublattices $A$ and $B$ such that 
there is no connection (i.e., bond) within the same sublattice. Namely, 
all nearest-neighbor sites $j^{\prime}$ of site $j \in A$ are $j^{\prime} \in B$ and vice versa. 
We also assume that the numbers $L_A$ and $L_B$ of sites in sublattices $A$ and $B$, respectively, 
are the same, i.e., $L_A = L_B = L/2$, where $L$ is the number of sites in the whole lattice and is assumed 
to be even.

$\hat{\bm S}_j = (\hat{S}^x_j, \hat{S}^{y}_j, \hat{S}^{z}_j)$ in Eq.~(\ref{eq:ham:ex}) denotes the spin operator of a mobile electron 
given by
\begin{equation}
  \hat{S}^{\mu}_j = \frac{1}{2} \hat{\bm c}^{\dagger}_j \mbox{\boldmath{$\sigma$}}_{\mu} \hat{\bm c}_j
\end{equation}
where
\begin{align}
  \hat{\bm c}^{\dagger}_j = (
  \begin{array}{cc}
    \hat{c}_{j\uparrow}^{\dagger} &
    \hat{c}_{j\downarrow}^{\dagger} \\
  \end{array}
  ), \\
  \hat{\bm c}_j = \left(
  \begin{array}{c}
    \hat{c}_{j\uparrow} \\
    \hat{c}_{j\downarrow} \\
  \end{array}
  \right), 
\end{align}
and $\mbox{\boldmath{$\sigma$}}_{\mu}$ ($\mu = x,y,z$) is the
$\mu$-component of the Pauli matrix.
$\hat{\bm M}_j = (\hat{M}_j^x, \hat{M}_j^y, \hat{M}_j^z)$ denotes the spin-$1/2$ operator for the localized spin at site $j$, which 
is coupled to the mobile electrons via the antiferromagnetic exchange interaction $J\,(>0)$.

In this study, we consider 
the half-filling case with the number of mobile electrons $N = L$. 
In this case, the ground state of the Kondo lattice model 
is insulating. 
However, the symmetry properties discussed in this section are not limited to the half-filling case
and can be easily extended to the case away from half filling.

\subsection{\label{sec:etaop}$\mbox{\boldmath{$\eta$}}$-pairing operators for Kondo lattice model}
C. N. Yang was the first who has noticed, in addition to the usual SU(2) rotational symmetry in the spin space, 
there exists an additional SU(2) pseudo-spin symmetry for the Hubbard model~\cite{PhysRevLett.63.2144}. 
A similar symmetry structure is also found in the Kondo lattice model~\cite{tsunetsugu1997}. 
The pseudo-spin symmetry is described by the $\eta$-pairing operators defined as 
\begin{align}
  \hat{\eta}_x = & \frac{1}{2}
  \sum_j 
  {\rm e}^{{\rm i} \phi_j} (
  \hat{c}_{j\uparrow}^{\dagger} \hat{c}_{j\downarrow}^{\dagger} + 
  \hat{c}_{j\downarrow} \hat{c}_{j\uparrow} ),  \label{eq:eta_x}\\
  \hat{\eta}_y = & \frac{1}{2{\rm i}}
  \sum_{j}
  {\rm e}^{{\rm i} \phi_j} (
  \hat{c}_{j\uparrow}^{\dagger} \hat{c}_{j\downarrow}^{\dagger} -
  \hat{c}_{j\downarrow} \hat{c}_{j\uparrow} ), \label{eq:eta_y}\\
  \hat{\eta}_z = & \frac{1}{2}
  \sum_{j} (
  \hat{c}_{j\uparrow}^{\dagger} \hat{c}_{j\uparrow} + 
  \hat{c}_{j\downarrow}^{\dagger} \hat{c}_{j\downarrow} - 1 ), \label{eq:eta_z}
\end{align}
where $\phi_j$ is a phase factor given by
\begin{equation}
  \phi_j = \left\{
  \begin{array}{lll}
  0   & (\text{mod }2\pi) & \text{for }j \in A \\
  \pi & (\text{mod }2\pi) & \text{for }j \in B \\
  \end{array}
  \right.
\end{equation}
and hence ${\rm e}^{{\rm i}\phi_j} = {\rm e}^{-{\rm i}\phi_j} = 1$ or $-1$.
The exact form of $\phi_j$ is determined once 
the geometry of the bipartite lattice is fixed. For example, 
for the two-dimensional square lattice,
$\phi_j = {\bm q} \cdot {\bm r}_j$ with ${\bm q} = (\pi,\pi)$, where
${\bm r}_j$ indicates the position of site $j$ in the lattice. 
It is easy to show that these $\eta$-pairing operators 
$\hat{\mbox{\boldmath{$\eta$}}} = (\hat{\eta}_x, \hat{\eta}_y, \hat{\eta}_z)$ 
satisfy the SU(2) commutation relations: 
\begin{equation}
  \left[ \hat{\eta}_{\mu}, \hat{\eta}_{\nu} \right] =
       {\rm i}\sum_\lambda \varepsilon_{\mu \nu \lambda} \hat{\eta}_{\lambda}
       \label{eq:eta:as:spin}
\end{equation}
for $\mu, \nu, \lambda = x,y,z$,
where $\varepsilon_{\mu \nu \lambda}$ is the Levi-Civita symbol.

We can also easily show that the $\eta$-pairing operators commute with the spin operators 
of mobile electrons, $\hat{\bm S}_j = (\hat{S}^x_j, \hat{S}^{y}_j, \hat{S}^{z}_j)$, 
i.e., 
\begin{equation}
[ \hat{\eta}_{\mu}, \hat{S}^{\mu}_j ] = 0, 
\end{equation}
as well as the kinetic part of the Hamiltonian, i.e., 
\begin{equation}
  [ \hat{\eta}_{\mu}, \hat{\mathcal{H}}_t ] = 0. 
\end{equation}
Since the $\eta$-pairing operators also commute with the localized spins 
$\hat{\bm M}_j = (\hat{M}_j^x, \hat{M}_j^y, \hat{M}_j^z)$, we find that 
the $\eta$-pairing operators commute with the Hamiltonian: 
\begin{equation}
[ \hat{\eta}_{\mu}, \hat{\mathcal{H}} ] = 0. \label{eq:comm:eta-ham}
\end{equation}
This implies that the Kondo lattice model is symmetric under the SU(2) pseudo-spin 
rotation, which is referred to as the $\eta$-SU(2) symmetry. Note that Eq.~(\ref{eq:comm:eta-ham}) 
is satisfied for all the components of $\eta$-pairing operators at any concentration of mobile electrons, 
while the corresponding commutation relations for the $x$ and $y$ components are fulfilled only 
at half filling for the Hubbard model~\cite{PhysRevLett.63.2144}. 

In the Kondo lattice model, the total spin operators of mobile electrons
\begin{align}
  \hat{S}_{\mu} = \sum_{j} \hat{S}^{\mu}_j
\end{align}
for $\mu = x,y,z$ do not commute with the Hamiltonian but
the total spin operators including the localized spin operators
\begin{equation}
\hat{S}_{\rm tot}^{\mu} = \sum_{j} ( \hat{S}^{\mu}_j + \hat{M}^{\mu}_j )
\end{equation}
commute with the Hamiltonian $\hat{\mathcal{H}}$. 
The $\eta$-pairing operators characterize the symmetry
related to the charge degrees of freedom, 
which is frozen for the localized spins in the Kondo lattice model.

Noticing that the total pseudo-spin operator squared that is defined as 
\begin{equation}
  \hat{\mbox{\boldmath{$\eta$}}}^2 = \hat{\eta}_x^2 + \hat{\eta}_y^2 + \hat{\eta}_z^2
\end{equation}
commutes with each component of the $\eta$-pairing operators, 
\begin{equation}
[ \hat{\eta}_{\mu}, \hat{\mbox{\boldmath{$\eta$}}}^2 ] = 0,  
\end{equation}
Eq.~(\ref{eq:comm:eta-ham}) suggests that we can block-diagonalize an eigenstate of $\hat{\mathcal{H}}$ by
quantum numbers $(\eta, \eta_z)$ for operators $\hat{\mbox{\boldmath{$\eta$}}}^2$ and $\hat{\eta}_z$. 
Let $\vert n, \eta, \eta_z \rangle$ be a simultaneous eigenstate 
for $\hat{\mathcal{H}}$, $ \hat{\mbox{\boldmath{$\eta$}}}^2$, and $\hat{\eta}_z$:
\begin{align}
  & \hat{\mathcal{H}} \vert n, \eta, \eta_z \rangle = E_{n \eta} \vert n, \eta, \eta_z \rangle, \\
  & \hat{\mbox{\boldmath{$\eta$}}}^2 \vert n, \eta, \eta_z \rangle =
  \eta (\eta + 1 ) \vert n, \eta, \eta_z \rangle, \\
  & \hat{\eta}_z \vert n, \eta, \eta_z \rangle = \eta_z \vert n, \eta, \eta_z \rangle. 
\end{align}
Note here that an energy eigenvalue $E_{n\eta}$ is independent of $\eta_z$, 
unlike the case of the Hubbard model~\cite{PhysRevLett.63.2144}. This is because of the fact that
\begin{equation}
  [ \hat{\eta}_{\pm}, \hat{\mathcal{H}} ] = 0,
\end{equation}
where 
\begin{equation}
  \hat{\eta}_{\pm} = \hat{\eta}_x \pm {\rm i} \hat{\eta}_y,  
\end{equation}
i.e., $\hat{\eta}_+ = \sum_j {\rm e}^{{\rm i} \phi_j} 
  \hat{c}_{j\uparrow}^{\dagger} \hat{c}_{j\downarrow}^{\dagger} $ and 
  $\hat{\eta}_- = \sum_j 
  {\rm e}^{-{\rm i} \phi_j}  
  \hat{c}_{j\downarrow} \hat{c}_{j\uparrow} $, and thus $\left[\hat{\eta}_+,\hat{\eta}_-\right]=2\hat{\eta}_z$. 

A simple example of simultaneous eigenstates 
is the vacuum state of mobile electrons, corresponding to $(\eta,\eta_z) = (L/2, -L/2)$. 
Here, the vacuum state can be represented as
\begin{equation}
  \vert {\rm vac} \rangle \equiv \vert 0 \rangle_f \otimes \vert \mbox{\boldmath{$\sigma$}} \rangle_S,
\end{equation}
where $\vert 0 \rangle_f$ indicates the vacuum of mobile electrons in the conduction band 
and $\vert \mbox{\boldmath{$\sigma$}} \rangle_S$ with
\begin{equation}
  \mbox{\boldmath{$\sigma$}} = \{ \sigma_1, \sigma_2, \cdots, \sigma_j, \cdots, \sigma_L \}
  \quad (\sigma_j = \pm 1/2)
\end{equation}
denotes a spin configuration of the localized spins.
Indeed, we can readily find that
\begin{align}
  & \hat{\mathcal{H}} \vert {\rm vac} \rangle = 0, \\
  & \hat{\mbox{\boldmath{$\eta$}}}^2
    \vert {\rm vac} \rangle =
  \frac{L}{2} \left( \frac{L}{2} + 1 \right) \vert {\rm vac} \rangle, \\
  & \hat{\eta}_z \vert {\rm vac} \rangle =
  - \frac{L}{2} \vert {\rm vac} \rangle.
\end{align}
Therefore, we can conclude that
\begin{equation}
  \vert {\rm vac} \rangle = \vert n, L/2, -L/2 \rangle,
\end{equation}
where $n$ can be used to label the states for different spin configurations $\mbox{\boldmath{$\sigma$}}$. 
Note that the vacuum states are macroscopically degenerate 
due to the spin configurations of the localized spins, and thus the localized spins behave paramagnetic with  
no effective interaction mediated via mobile electrons.

By applying the $\hat{\eta}_+$ operator sequentially onto $\vert {\rm vac} \rangle$,
we can obtain an energy eigenstate for the different number of mobile electrons:  
\begin{equation}
  (\hat{\eta}_+)^{N/2} \vert {\rm vac} \rangle \propto
  \vert n, L/2, -L/2 + N/2 \rangle,
  \label{eq:yangstate}
\end{equation}
which contains $N$ mobile electrons in the conduction band. 
Since the energy is independent of $\eta_z$, the states with $\eta = L/2$
are degenerate macroscopically and paramagnetic.
The state given in Eq.~(\ref{eq:yangstate})
can also be obtained by applying the number projection to a BCS-type wave function:
\begin{equation}
  \vert n, L/2, -L/2 + N/2 \rangle \propto \hat{\mathcal{P}}_{N} \vert {\rm BCS} \rangle 
\end{equation}
with
\begin{equation}
  \vert {\rm BCS} \rangle = \exp \left[ \hat{\eta}_+ \right] \vert 0 \rangle_f \otimes \vert \mbox{\boldmath{$\sigma$}} \rangle_S 
\end{equation}
where $\hat{\mathcal{P}}_N$ denotes the projection operator onto the subspace with $N$ mobile electrons. 
It is now clear that the state given in Eq.~(\ref{eq:yangstate}) exhibits the off-diagonal long-range order 
characterized by the pair correlation function $P_{\eta}$ given by
\begin{equation}
    P_{\eta} = \frac{1}{L} \sum_{j, j^{\prime}}
    {\rm e}^{{\rm i} ( \phi_j - \phi_{j^{\prime}} )}
  \langle
  \hat{c}_{j\uparrow}^{\dagger} \hat{c}_{j\downarrow}^{\dagger}
  \hat{c}_{j^{\prime}\downarrow} \hat{c}_{j^{\prime}\uparrow}
  \rangle 
  = 
  \frac{1}{L} \langle \hat{\eta}_+ \hat{\eta}_- \rangle
\end{equation}
where $\langle \cdots \rangle = \langle \psi \vert \cdots \vert \psi \rangle$
indicates the expectation value for a given state $\vert \psi \rangle$.
We can easily verify that, for the state given in Eq.~(\ref{eq:yangstate}),
\begin{equation}
P_{\eta} = \frac{N (2L-N+2)}{4L}.
\end{equation}
This implies that $P_{\eta} \propto L$ provided that $N \propto L$ for $L \to \infty$,
suggesting the long-range ordering. 
Finally, we note that these states are not the ground state but energetically higher 
for a given $\eta_z$. 
Thereby, these properties are usually masked by the thermal average~\cite{kaneko2020}.

\subsection{\label{sec:tensor}Tensor operators}

Next, let us explain the relation between the $\eta$-pairing operators and the current operator.
To this end, we introduce the following set of operators:
\begin{align}
  \hat{\mathcal{J}}^{(0)}_{\alpha} = & - {\rm i} t \sum_{\langle j, j^{\prime} \rangle} \sum_{\sigma = \uparrow,\downarrow}
  d^{\alpha}_{jj^{\prime}}
  (
  \hat{c}_{j\sigma}^{\dagger} \hat{c}_{j^{\prime}\sigma} -
  \hat{c}_{j^{\prime}\sigma}^{\dagger} \hat{c}_{j\sigma}
  ), \label{eq:current:0} \\
  \hat{\mathcal{J}}^{(1)}_{\alpha} = & - \sqrt{2} {\rm i} t
  \sum_{\langle j,j^{\prime}\rangle}
  {\rm e}^{{\rm i}\phi_{j}}
  d^{\alpha}_{jj^{\prime}}
  (
  \hat{c}_{j\uparrow}^{\dagger} \hat{c}_{j^{\prime}\downarrow}^{\dagger} + 
  \hat{c}_{j^{\prime}\uparrow}^{\dagger} \hat{c}_{j\downarrow}^{\dagger}
  ), \label{eq:current:p}\\
  \hat{\mathcal{J}}^{(-1)}_{\alpha} = & - \sqrt{2} {\rm i} t
  \sum_{\langle j,j^{\prime} \rangle}
  {\rm e}^{{\rm i}\phi_{j}} d^{\alpha}_{jj^{\prime}}
  (
  \hat{c}_{j\downarrow} \hat{c}_{j^{\prime}\uparrow} +
  \hat{c}_{j^{\prime}\downarrow} \hat{c}_{j\uparrow}
  ). \label{eq:current:m}
\end{align}
Here, $d^{\alpha}_{jj^{\prime}}\,{(=-d^{\alpha}_{j^{\prime}j})}$ is a scalar and depends on 
sites $j$ and $j^{\prime}$. $d^{\alpha}_{jj^{\prime}}$ can be chosen arbitrary 
as long as sites $j$ and $j^{\prime}$ belong to different sublattices of a bipartite lattice, 
implying that ${\rm e}^{{\rm i}\phi_{j}}=-{\rm e}^{{\rm i}\phi_{j^{\prime}}}$.
A practical choice of $d^{\alpha}_{jj^{\prime}}$ is 
\begin{equation}
  d^{\alpha}_{jj^{\prime}} =  ( {\bm r}_j - {\bm r}_{j^{\prime}} ) \cdot {\bm e}_{\alpha},
  \label{eq:dajj}
\end{equation}
where ${\bm r}_j$ indicates the position of site $j$ 
and ${\bm e}_{\alpha}$ denotes the unit vector
pointing to an arbitrary direction $\alpha$.
In this case, $\hat{\mathcal{J}}^{(0)}_{\alpha}$ corresponds to 
the current operator for the $\alpha$ direction. 
Note that we define $\hat{\mathcal{J}}_{\alpha}^{(-1)}$ so as to satisfy
$(\hat{\mathcal{J}}_{\alpha}^{(+1)})^{\dagger} = - \hat{\mathcal{J}}_{\alpha}^{(-1)}$. 
In this case, we can show that these three operators satisfy the following commutation relations:
\begin{equation}
  \begin{split}
  & [ \hat{\eta}_{\pm}, \hat{\mathcal{J}}_{\alpha}^{(q)} ] = \sqrt{(1\mp q)(1 \pm q+1)} \hat{\mathcal{J}}^{(q \pm 1)}_{\alpha}, \\
  & [ \hat{\eta}_{z}, \hat{\mathcal{J}}_{\alpha}^{(q)} ] = q \hat{\mathcal{J}}_{\alpha}^{(q)}
  \end{split}
  \label{eq:tensor:current}
\end{equation}
for $q=-1,0,1$.
These relations in Eq.~(\ref{eq:tensor:current}) suggest that
the set of operators $\hat{\mathcal{J}}^{(q)}_{\alpha}$ is  
a rank-1 tensor operator for the pseudo-spin operators $\hat{\eta}_{\mu}$ ($\mu =x,y,z$).

We can use the Wigner-Eckert theorem to evaluate a matrix element of a tensor operator between 
two states $ \vert  n, \eta, \eta_z\rangle$ and $\vert n^{\prime}, \eta^{\prime},\eta^{\prime}_z \rangle$~\cite{jjsakurai}.
The theory states that, given the $q$-th component of a tensor operator $\hat{\mathcal{T}}_{kq}$ of rank $k$,
there exists a constant $\langle n \eta \vert\vert \hat{\mathcal{T}}_k \vert \vert n^{\prime} \eta^{\prime} \rangle$, 
referred to as a reduced matrix element,  
such that for all $\eta_z$, $\eta_z^{\prime}$, and $q$,
\begin{equation}
  \langle n,\eta,\eta_z \vert \hat{\mathcal{T}}_{kq} \vert n^{\prime}, \eta^{\prime},\eta_z^{\prime} \rangle
  = \langle \eta^{\prime} \eta_z^{\prime} k q \vert \eta \eta_z \rangle
  \langle n \eta \vert\vert \hat{\mathcal{T}}_{k} \vert \vert n^{\prime} \eta^{\prime} \rangle
  \label{eq:wet}
\end{equation}
where $\langle \eta^{\prime} \eta^{\prime}_z k q \vert \eta \eta_z \rangle$
is the Clebsch-Gordan coefficient 
and $\langle n \eta \vert\vert \hat{\mathcal{T}}_k \vert \vert n^{\prime} \eta^{\prime} \rangle$ is independent of 
$\eta_z$, $\eta^{\prime}_z$, and $q$. 
Therefore, there is a finite matrix element for the current operator $\hat{\mathcal{J}}^{(0)}_{\alpha}$ 
only when $\eta^{\prime}=\eta\pm1$ and $\eta$. 
As we shall show below, this selection rule is essential 
when we discuss the photoexcitation.

Next, let us introduce the following kinetic energy operator 
in the $\alpha$ direction: 
\begin{equation}
  \hat{\mathcal{K}}^{(0)}_{\alpha} = - t \sum_{\langle j,j^{\prime} \rangle}
  b^{\alpha}_{jj^{\prime}}
  (
  \hat{c}_{j\sigma}^{\dagger} \hat{c}_{j^{\prime}\sigma} +
  \hat{c}_{j^{\prime}\sigma}^{\dagger} \hat{c}_{j\sigma}
  ),
  \label{eq:kinetic:0}
\end{equation}
where $b^{\alpha}_{jj^{\prime}}\,(=b^{\alpha}_{j^{\prime}j})$ is a scalar and 
depends on sites $j$ and $j^{\prime}$ belonging to different sublattices of a bipartite lattice. 
We can readily show that $\hat{\mathcal{K}}^{(0)}_{\alpha}$ commutes with
the $\eta$-pairing operators: 
\begin{equation}
  \begin{split}
    [ \hat{\eta}_{\pm}, \hat{\mathcal{K}}^{(0)}_{\alpha} ] = 0, \\
    [ \hat{\eta}_{z}, \hat{\mathcal{K}}^{(0)}_{\alpha} ] = 0, 
  \end{split}
  \label{eq:tensor:kinetic}
\end{equation}
suggesting that $\hat{\mathcal{K}}_{\alpha}^{(0)}$ is a rank-0 tensor operator 
for the pseudo-spin operators $\hat{\eta}_{\mu}$ ($\mu =x,y,z$).
The kinetic term $\hat{\mathcal{H}}_t$ in the Hamiltonian $\hat{\mathcal{H}}$ 
corresponds to the case when $b^{\alpha}_{jj^{\prime}}=1$ and thus  
it is a tensor operator of rank 0.

\subsection{\label{sec:simulation}Time-dependent electric field}

We introduce a time-dependent external field
via the Peierls substitution 
by replacing $\hat{\mathcal{H}}_t$ in Eq.~(\ref{eq:ham:hop}) with
$\hat{\mathcal{H}}_t(\tau)$ given by
\begin{equation}
  \hat{\mathcal{H}}_t (\tau) = - t \sum_{\langle j, j^{\prime} \rangle}
  \sum_{\sigma = \uparrow,\downarrow} 
  ( {\rm e}^{{\rm i}A_{jj^{\prime}}(\tau)} \hat{c}_{j\sigma}^{\dagger} \hat{c}_{j^{\prime}\sigma}
  + {\rm e}^{{\rm i}A_{j^{\prime}j}(\tau)} \hat{c}_{j^{\prime}\sigma}^{\dagger} \hat{c}_{j\sigma} )
\end{equation}
where $A_{jj^{\prime}}(\tau)$ is the vector potential
as a function of time $\tau$, 
\begin{equation}
  A_{jj^{\prime}}(\tau) = A(\tau)  ({\bm r}_j - {\bm r}_{j^{\prime}})\cdot {\bm e}_{\alpha}. 
\end{equation}
For simplicity, the light velocity, the elementary charge, the Planck constant,
and the lattice constant are set to 1.
In this study, we consider the pump pulse given by 
\begin{equation}
  A(\tau) = A_0 {\rm e}^{-(\tau - \tau_c)^2/(2 \tau_w^2)}
  \cos [ \omega_p ( \tau - \tau_c) ]
\end{equation}
with the amplitude $A_0$, frequency $\omega_p\, (>0)$,
and pulse width $\tau_w$ centered at time $\tau_c$. 
This implies that a time-dependent electric field is applied along the $\alpha$ direction. 

\subsection{\label{sec:perturb}Time-dependent perturbation theory}

It is highly instructive to analyze the effect of the time-dependent external field introduced above by using the time-dependent 
perturbation theory. For this purpose, we should first notice that the Hamiltonian $\hat{\mathcal{H}}(\tau)$ 
with the time-dependent external field can be decomposed as 
\begin{equation}
  \begin{split}
  \hat{\mathcal{H}}(\tau) = & \hat{\mathcal{H}}_t(\tau) + \hat{\mathcal{H}}_J \\
  = & \hat{\mathcal{H}} + 
  \hat{\mathcal{K}}^{(0)}_{\alpha}(\tau) +
  \hat{\mathcal{J}}^{(0)}_{\alpha}(\tau)
  \label{eq:ham:tdep}
  \end{split}
\end{equation}
where the two classes of perturbation terms are 
\begin{equation}
\hat{\mathcal{K}}^{(0)}_{\alpha}(\tau) = - t \sum_{\langle j,j^{\prime} \rangle} \sum_{\sigma = \uparrow,\downarrow}
  (\cos  A_{jj^{\prime}}(\tau)  - 1 )
  (
  \hat{c}_{j\sigma}^{\dagger} \hat{c}_{j^{\prime}\sigma} +
  \hat{c}_{j^{\prime}\sigma}^{\dagger} \hat{c}_{j\sigma}
  )
  \label{eq:kinetic:t}
\end{equation}
and 
\begin{equation}
  \hat{\mathcal{J}}^{(0)}_{\alpha}(\tau) = - {\rm i} t \sum_{\langle j,j^{\prime} \rangle} \sum_{\sigma = \uparrow,\downarrow}
  \sin  A_{jj^{\prime}}(\tau) 
  (
  \hat{c}_{j\sigma}^{\dagger} \hat{c}_{j^{\prime}\sigma} -
  \hat{c}_{j^{\prime}\sigma}^{\dagger} \hat{c}_{j\sigma}
  ).
  \label{eq:current:t}
\end{equation}
Since $\cos  A_{jj^{\prime}}(\tau) $ [$\sin  A_{jj^{\prime}}(\tau) $] is even (odd) under the exchange of $j$ and $j^{\prime}$,
$\hat{\mathcal{K}}^{(0)}_{\alpha}(\tau)$ is a form of the kinetic energy operator defined in Eq.~(\ref{eq:kinetic:0}) and 
thus a tensor operator of rank 0, while $\hat{\mathcal{J}}^{(0)}_{\alpha}(\tau)$ is a form of one of 
the three operators introduced in Eqs.~(\ref{eq:current:0})--(\ref{eq:current:m}) and thus a tensor operator of rank 1. 
This implies that the time-dependent external field can excite a state to other states with $\eta$ different at most by 1 
in each order of the perturbation.

To explore this more explicitly, let us analyze the effect of the time-dependent external field using 
the time-dependent perturbation theory in the limit of $\tau_{w} \to \infty$. Here we also set $\tau_c = 0$, for simplicity. 
The similar analysis has been described briefly in Supplementary Information of Ref.~\cite{PhysRevLett.122.077002}. 
In the limit of $\tau_{w} \to \infty$, 
we can simply Fourier expand the $\tau$-dependent parts of the perturbations  
$\hat{\mathcal{K}}^{(0)}_{\alpha}(\tau)$ and $\hat{\mathcal{J}}^{(0)}_{\alpha}(\tau)$ as  
\begin{equation}
  \begin{split}
  & \cos  A_{jj^{\prime}}(\tau) - 1 = \sum_{n=-\infty}^{\infty} {\rm e}^{-{\rm i}2n\omega_p \tau} f_{jj^{\prime}}^{(2n)}, \\
  & \sin  A_{jj^{\prime}}(\tau) = \sum_{n=-\infty}^{\infty} {\rm e}^{-{\rm i}(2n+1)\omega_p \tau} g_{jj^{\prime}}^{(2n+1)}
  \end{split}
\end{equation}
where $n$ is integer, and $f_{jj^{\prime}}^{(n)}$ and $g_{jj^{\prime}}^{(n)}$ are the Fourier coefficients 
given by
\begin{equation}
  \begin{split}
    & f_{jj^{\prime}}^{(n)} = \frac{1}{2\pi} \int_{-\infty}^{\infty} {\rm d}\tau \left( \cos  A_{jj^{\prime}}(\tau)  - 1 \right) {\rm e}^{{\rm i}n \omega_p \tau}, \\
    & g_{jj^{\prime}}^{(n)} = \frac{1}{2\pi} \int_{-\infty}^{\infty} {\rm d}\tau \sin  A_{jj^{\prime}}(\tau)  {\rm e}^{{\rm i}n \omega_p \tau}.
  \end{split}
  \label{eq:fg}
\end{equation}
Note that the integral can be performed explicitly and
the results are represented by using the Bessel functions $J_n(x)$~\cite{PhysRevB.94.174503} as follows:  
\begin{align}
& f_{jj^{\prime}}^{(2n)} = (-1)^{n} ( J_{2n} (A_0d_{jj^{\prime}}^{\alpha}) - \delta_{n0} ), \\
& g_{jj^{\prime}}^{(2n+1)} = (-1)^{n} J_{2n+1} (A_0 d_{jj^{\prime}}^{\alpha}), 
\end{align}
with $d_{jj^{\prime}}^{\alpha}$ in Eq.~(\ref{eq:dajj})
and $f_{jj^{\prime}}^{(2n+1)} = g_{jj^{\prime}}^{(2n)} = 0$.
Using these Fourier expansions, we obtain
\begin{equation}
  \hat{\mathcal{V}}(\tau) = \hat{\mathcal{K}}^{(0)}_{\alpha}(\tau) + \hat{\mathcal{J}}^{(0)}_{\alpha}(\tau)
   = \sum_{n=-\infty}^{\infty} {\rm e}^{-{\rm i}n\omega_p \tau} \hat{\mathcal{V}}_n, 
   \label{eq:v_t}
\end{equation}
where
\begin{align}
  \hat{\mathcal{V}}_{2n}   = & - t \sum_{\langle j,j^{\prime} \rangle} \sum_{\sigma = \uparrow,\downarrow} f_{jj^{\prime}}^{(2n)} 
  (\hat{c}_{j\sigma}^{\dagger} \hat{c}_{j^{\prime}\sigma} + \hat{c}_{j^{\prime}\sigma}^{\dagger} \hat{c}_{j\sigma} ), \\
  \hat{\mathcal{V}}_{2n+1} = & - {\rm i} t \sum_{\langle j,j^{\prime} \rangle} \sum_{\sigma = \uparrow,\downarrow} g_{jj^{\prime}}^{(2n+1)} 
  (\hat{c}_{j\sigma}^{\dagger} \hat{c}_{j^{\prime}\sigma} - \hat{c}_{j^{\prime}\sigma}^{\dagger} \hat{c}_{j\sigma} ). 
\end{align}
Since $f_{jj^{\prime}}^{(2n)} = f_{j^{\prime}j}^{(2n)} $ and $g_{jj^{\prime}}^{(2n+1)} = -g_{j^{\prime}j}^{(2n+1)}$, 
we find that the even terms $\hat{\mathcal{V}}_{2n} $ are rank-0 tensor operators and 
the odd terms $\hat{\mathcal{V}}_{2n+1}$ are rank-1 tensor operators [see Eqs.~(\ref{eq:current:0}) and (\ref{eq:kinetic:0})].

The time-dependent wave function $\vert \psi (\tau) \rangle$
is generally expanded in terms of the eigenstates
$\vert \psi_m \rangle$ of the unperturbed Hamiltonian $\hat{\mathcal{H}}$ with the energies $E_m$: 
\begin{equation}
\vert \psi (\tau) \rangle = 2 \pi {\rm i} \sum_{m} c_{m} (\tau) \vert \psi_m \rangle, 
\end{equation}
where $m=0,1,2,\dots$ and $\vert \psi_0 \rangle$ corresponds to the ground state of $\hat{\mathcal{H}}$ 
with $E_0< E_1\le E_2\le\dots$, assuming that the ground state is not degenerate. 
In the perturbation theory,
the coefficient $c_{m}(\tau)$ is expanded by the order $k$ of the perturbation:
\begin{equation}
  c_{m}(\tau) = \sum_{k=0}^{\infty} c_{m}^{(k)} (\tau)  
\end{equation}
with the initial condition that $\vert \psi (\tau=-\infty) \rangle =\vert \psi_0 \rangle$, i.e. 
$c_m^{(k=0)}(\tau=-\infty)=\frac{1}{2 \pi {\rm i}}\delta_{m0}$.

According to the time-dependent perturbation theory, 
$c_m^{(k)}(\tau)$ is given as  
\begin{widetext}
\begin{align}
c_m^{(k)}(\tau) =
\frac{( - {\rm i} )^k}{2 \pi {\rm i}} 
\int_{-\infty}^{\tau}  \! d \tau_k 
 \cdots \!
\int_{-\infty}^{\tau_{3}}  \! d \tau_2 
\int_{-\infty}^{\tau_{2}}  \! d \tau_1
\sum_{m_{k-1}}  
 \cdots \sum_{m_2}   \sum_{m_1}  
\braket{ \psi_m | \hat{\mathcal{V}}_I (\tau_k) | \psi_{m_{k-1}} } 
\cdots   \!
\braket{ \psi_{m_{2}} | \hat{\mathcal{V}}_I (\tau_2) | \psi_{m_{1}} }
\braket{ \psi_{m_{1}} | \hat{\mathcal{V}}_I (\tau_1) | \psi_{0} }, 
\label{eq:tdpt}
\end{align}
\end{widetext}
where $\hat{\mathcal{V}}_I(t) = e^{{\rm i}\hat{\mathcal{H}}t} \hat{\mathcal{V}}(t) e^{- {\rm i}\hat{\mathcal{H}}t}$.
Because of Eq.~(\ref{eq:v_t}), we can find the explicit $\tau$-dependance of each matrix element as 
\begin{align}
\braket{ \psi_m | \hat{\mathcal{V}}_I(t)| \psi_{m'} } 
&= \sum_{n=-\infty}^\infty {\rm e}^{{\rm i}(E_m-E_{m'}-n\omega_p)\tau} \mathcal{V}^{(n)}_{m,m'}
\label{HM_etaSR_eq94} 
\end{align}
with 
\begin{align}
\mathcal{V}_{m,m^{\prime}}^{(n)} = \langle \psi_m \vert \hat{\mathcal{V}}_n \vert \psi_{m^{\prime}} \rangle.  
\label{eq:mtel}
\end{align}
Therefore, taking $\tau\to\infty$, we obtain that 
\begin{equation}
c_m^{(k=1)}(\infty) =  -  \sum_{n=-\infty}^\infty 
\mathcal{V}_{m, 0}^{(n)} 
 \delta ( E_m - E_0 -  n \omega_p )
\label{eq:perturbation0}
\end{equation}
and for $k>1$
\begin{widetext}
\begin{eqnarray}
c_m^{(k)}(\infty) &=&  (-1)^k \sum_{n_k=-\infty}^\infty \cdots \sum_{n_2=-\infty}^\infty \sum_{n_1=-\infty}^\infty 
\sum_{m_{k-1}} \cdots \sum_{m_2} \sum_{m_1} 
\mathcal{V}_{m, m_{k-1}}^{(n_k)} \cdots 
\mathcal{V}_{m_2, m_1}^{(n_2)} 
\mathcal{V}_{m_1, 0}^{(n_1)} 
\prod_{k^{\prime}=1}^{k-1} 
\frac{1}{E_{m_{k^{\prime}}}-E_0 - \left(\sum_{\ell=1}^{k^{\prime}} n_{\ell}\right) \omega_p - i \delta}
\nonumber \\
& &\quad\quad \times \delta \left( E_m - E_0 - \left(\sum_{\ell=1}^k n_{\ell}\right) \omega_p \right),
\label{eq:perturbation}
\end{eqnarray}
\end{widetext}
where $\delta \to 0^+$ is a convergence factor.

\begin{figure}[htbp]
  \includegraphics[width=\hsize]{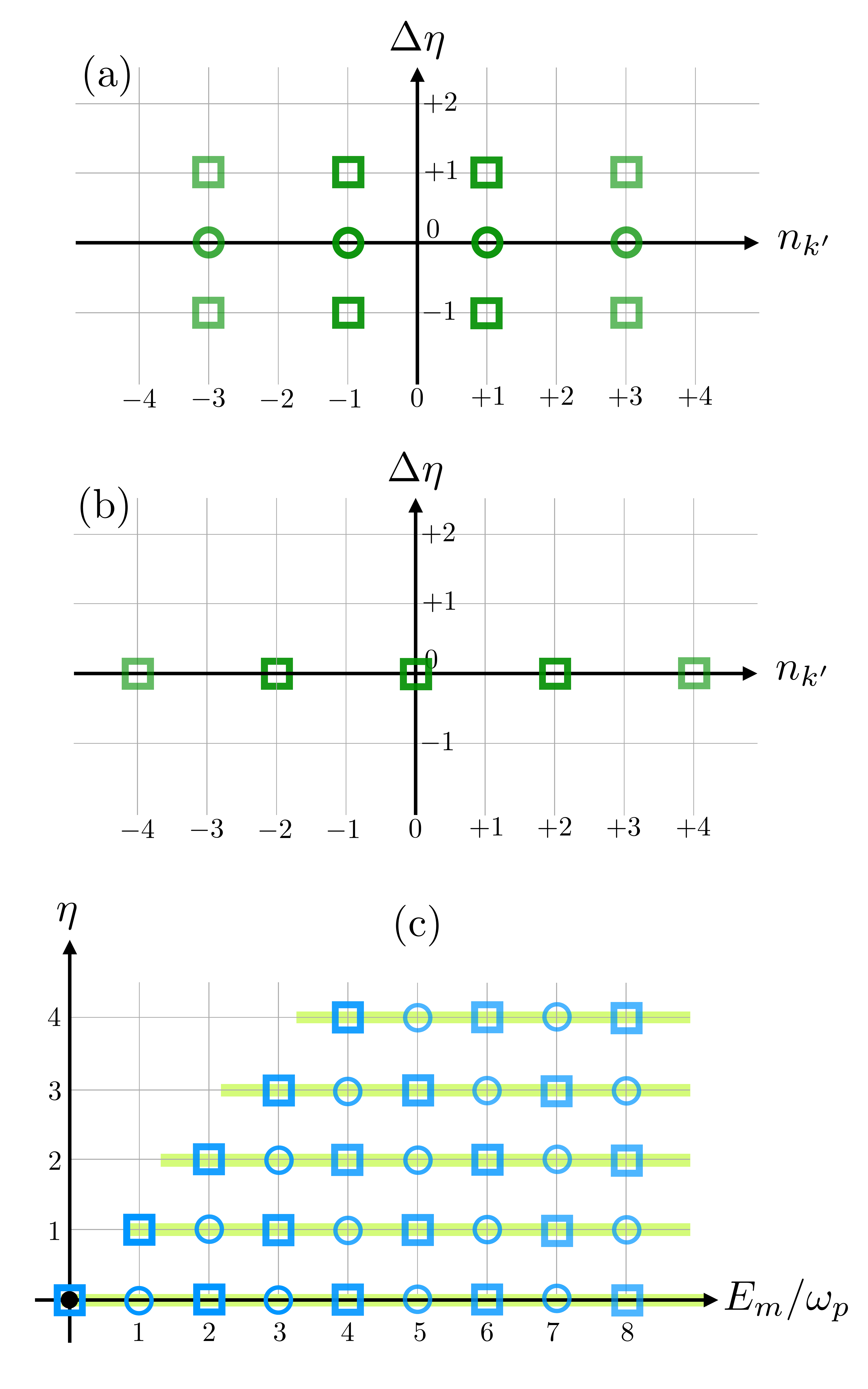}
  \caption{
  Schematic figures of possible transition processes by (a) a rank-1 tensor operator and
  (b) a rank-0 tensor operator, and (c) possible distribution of eigenstates
  for the final state. In (a) and (b), the vertical axis is the quantum number difference $\Delta\eta$ 
  and the horizontal axis is the energy difference $\Delta E$ in unit of $\omega_p$ (see the main text). 
  Open green squares indicate transitions allowed in general while open green circles indicate transitions allowed 
  only away from half filling. 
  In (c), the vertical axis is the value of $\eta$ and the horizontal axis is the energy $E_m$ 
  (in unit of $\omega_p$) of eigenstates distributed in the final state.
  Here, we assume the energy distribution of the unperturbed Hamiltonian $\hat{\mathcal{H}}$
  as indicated by light-green shaded bars. 
  The initial ground state is indicated by a black solid circle. 
  Open blue squares indicate eigenstates allowed in general while open blue circles indicate eigenstates allowed 
  only away from half filling. 
  }
  \label{fig:perturb:rule}
\end{figure}

Eq.~(\ref{eq:tdpt}) suggests that 
the transition from the initial state $\vert \psi_0 \rangle$ to the final state
$\vert \psi_{m} \rangle$ occurs via the intermediate states $\vert \psi_{m_{k'}}\rangle$ 
with $k^{\prime}=1,2,\cdots,k-1$. 
These intermediate states (and also the final state) are generated 
by applying the perturbations sequentially  represented by $\mathcal{V}_{m_{k'},m_{k'-1}}^{(n_{k'})}$ in Eq.~(\ref{eq:perturbation}), i.e,   
either by the rank-0 tensor operators when $n_{k'}$ is even or by the rank-1 tensor operators when $n_{k'}$ is odd. 
Therefore, this forces the selection rule for the transition between the two intermediate states $\vert \psi_{m_{k'-1}}\rangle$ 
and $\vert \psi_{m_{k'}}\rangle$: $\Delta\eta=\eta_{m_{k'}} - \eta_{m_{k'-1}}=0$ when $n_{k'}$ is even and 
$\Delta\eta=\pm1$ or $0$ when $n_{k'}$ is odd, where $\eta_{m_{k'}}(\eta_{m_{k'}}+1)$ is the eigenvalue of 
$\hat{\mbox{\boldmath{$\eta$}}}^2$ for $\vert \psi_{m_{k'}}\rangle$. 
The denominators in Eq.~(\ref{eq:perturbation}) 
suggests that the $k'$-th intermediate state $\vert \psi_{m_{k'}}\rangle$ contributes most when 
\begin{equation}
  E_{m_{k^{\prime}}} = E_0 + \left( \sum_{l=1}^{k^{\prime}} n_l \right) \omega_p
  = E_{m_{k^{\prime}-1}} +  n_{k^{\prime}} \omega_p,
\end{equation}
implying that the energy difference $\Delta E$ between the two intermediate states $\vert \psi_{m_{k'-1}}\rangle$ 
and $\vert \psi_{m_{k'}}\rangle$ is $\Delta E = E_{m_{k^{\prime}}} - E_{m_{k^{\prime}-1}} = n_{k^{\prime}} \omega_p$. 
This is indeed the energy conservation condition obtained by the first order perturbation theory with taking $\vert \psi_{m_{k'-1}}\rangle$ 
as the initial state [see Eq.~(\ref{eq:perturbation0})]. 
These rules are schematically summarized in Figs.~\ref{fig:perturb:rule}(a) and \ref{fig:perturb:rule}(b). 

The delta function in Eq.~(\ref{eq:perturbation}) 
determines the final state energy $E_{m}$ exactly as  
\begin{equation}
  E_{m} = E_0 + \left( \sum_{l=1}^k n_l \right) \omega_p.
\end{equation}
Therefore, the final state energy $E_{m}$ is larger than the initial ground state energy 
$E_0$ by (integer)$\times \omega_p$, also implying that  $\sum_{l=1}^k n_l\ge0$.
Since the final state is generated by applying the perturbations $k$ times through exciting 
the $k-1$ intermediate states, the same selection rule described above is also set to the final state and 
is schematically illustrated in Fig.~\ref{fig:perturb:rule}(c).

At half filling when $\eta_z = 0$, the selection rule is more characteristic because there is an additional rule
such that 
\begin{equation}
\langle n, \eta, 0 \vert \hat{\mathcal{V}}_{2n+1} \vert n^{\prime}, \eta,0 \rangle = 0
\end{equation}
because the Clebsch-Gordan coefficient in Eq.~(\ref{eq:wet}) is 
\begin{equation}
  \langle \eta 0 1 0 \vert \eta 0 \rangle = 0,
\end{equation}
suggesting that the transitions 
between two states with the same $\eta$ value by 
the rank-1 tensor operators $\hat{\mathcal{V}}_{2n+1}$ are prohibited [also see Fig.~\ref{fig:perturb:rule}(a)]. 
An important consequence of this is that the final state with 
an odd (even) value of $\eta$ is excited at the excitation energy (odd (even) integer)$\times\omega_p$, as schematically 
shown in Fig.~\ref{fig:perturb:rule}(c). 
This is because the final state with an odd (even) value of $\eta$ can be excited only by involving the rank-1 tensor operators 
odd (even) times.

\section{\label{sec:result}Numerical Calculations}

We first describe briefly the numerical setting of the calculations and numerical techniques used here, 
followed by the numerical results. 

\subsection{Numerical setting and techniques}

In what follows, we consider the one-dimensional (1D) periodic lattice composed of $L$ sites 
with the antiferromagnetic exchange interaction. 
In this case, the 1D Kondo lattice model is described by the following Hamiltonian: 
\begin{equation}
  \begin{split}
    \hat{\mathcal{H}}(\tau) = & \hat{\mathcal{H}}_t(\tau) + \hat{\mathcal{H}}_J, \\
    \hat{\mathcal{H}}_t(\tau) = & - t \sum_{j=1}^L \sum_{\sigma = \uparrow,\downarrow} 
    ( {\rm e}^{-{\rm i} A(\tau)} \hat{c}_{j\sigma}^{\dagger} \hat{c}_{j+1\sigma} + {\rm e}^{{\rm i}A(\tau)} \hat{c}_{j+1\sigma}^{\dagger} \hat{c}_{j\sigma} ), \\
    \hat{\mathcal{H}}_J = & J \sum_{j=1}^L \hat{\bm S}_j \cdot \hat{\bm M}_j,
  \end{split}
  \label{eq:1dkm}
\end{equation}
with $J>0$ and  $\hat{c}_{L+1\sigma} = \hat{c}_{1\sigma}$. 
At half filling ($N=L$),
the ground state of this model with $A(\tau)=0$
is an insulating state where both the spin and charge gaps open for any
$J/t$~\cite{PhysRevB.46.3175,PhysRevLett.71.3866,PhysRevLett.72.1048,doi:10.1143/JPSJ.66.2157}.
The presence of the spin gap is attributed to the local singlet formation 
between the mobile electron and the localized spin via the finite exchange interaction $J$. 
Note that the finite spin gap suggests the finite correlation length of
the antiferromagnetic correlation.
This is sharp contrast to the two-dimensional case where
there occurs the continuous quantum phase transition
between the antiferromagnetically ordered and spin-gapped insulating phases with increasing $J/t$~\cite{PhysRevLett.83.796}.
We however note that although the spin gap is finite, for small to moderate strength of the exchange interaction $J$,
the antiferromagnetic correlation is dominant as 
compared to other correlations because of the small spin gap~\cite{doi:10.1143/JPSJ.66.2157}.

We employ the exact diagonalization technique to perform the time-dependent simulation.
The initial state $\vert \psi (\tau=0) \rangle$ is set to be the ground state
of the Hamiltonian given in Eq.~(\ref{eq:1dkm}) with $A(\tau)=0$.
We obtain $\vert \psi (\tau=0) \rangle$ by using the standard Lanczos technique. 
We then calculate the time-evolved state $\vert \psi (\tau) \rangle$
by applying the time-evolution operator with the small time step $\delta \tau$ sequentially:
\begin{equation}
  \vert \psi (\tau + \delta \tau ) \rangle = {\rm e}^{-{\rm i}\hat{\mathcal{H}}(\tau) \delta \tau}
  \vert \psi (\tau) \rangle. 
  \label{eq:time-evolution}
\end{equation}
To deal with the exponential form of the time-evolution operator, we simply use the Taylor expansion:
\begin{equation}
  \vert \psi (\tau + \delta \tau) \rangle =
  \sum_{k=0}^{K} \vert v_k \rangle
\end{equation}
with
\begin{equation}
  \begin{split}
    & \vert v_0 \rangle = \vert \psi (\tau) \rangle, \\
  & \vert v_k \rangle = -\frac{{\rm i}\delta \tau}{k} \hat{\mathcal{H}}(\tau) \vert v_{k-1} \rangle\,\,\, \text{ for }\,k\geq 1. \\
  \end{split}
  \label{eq:teite}
\end{equation}
Note that, since the Hamiltonian is time-dependent,
one has to take the time step $\delta \tau$ small enough to reduce the systematic error, 
for which the Taylor expansion converges rather quickly.  
We set $\delta \tau = 0.01/t$ and determine 
$K$ flexibly so as to satisfy 
$\langle v_{K} \vert v_{K} \rangle < 10^{-12}$. 
The results shown below are for $L=8$ and $J=t$ at half filling.

\subsection{Time evolution of correlation functions}

\begin{figure}[htbp]
  \includegraphics[width=\hsize]{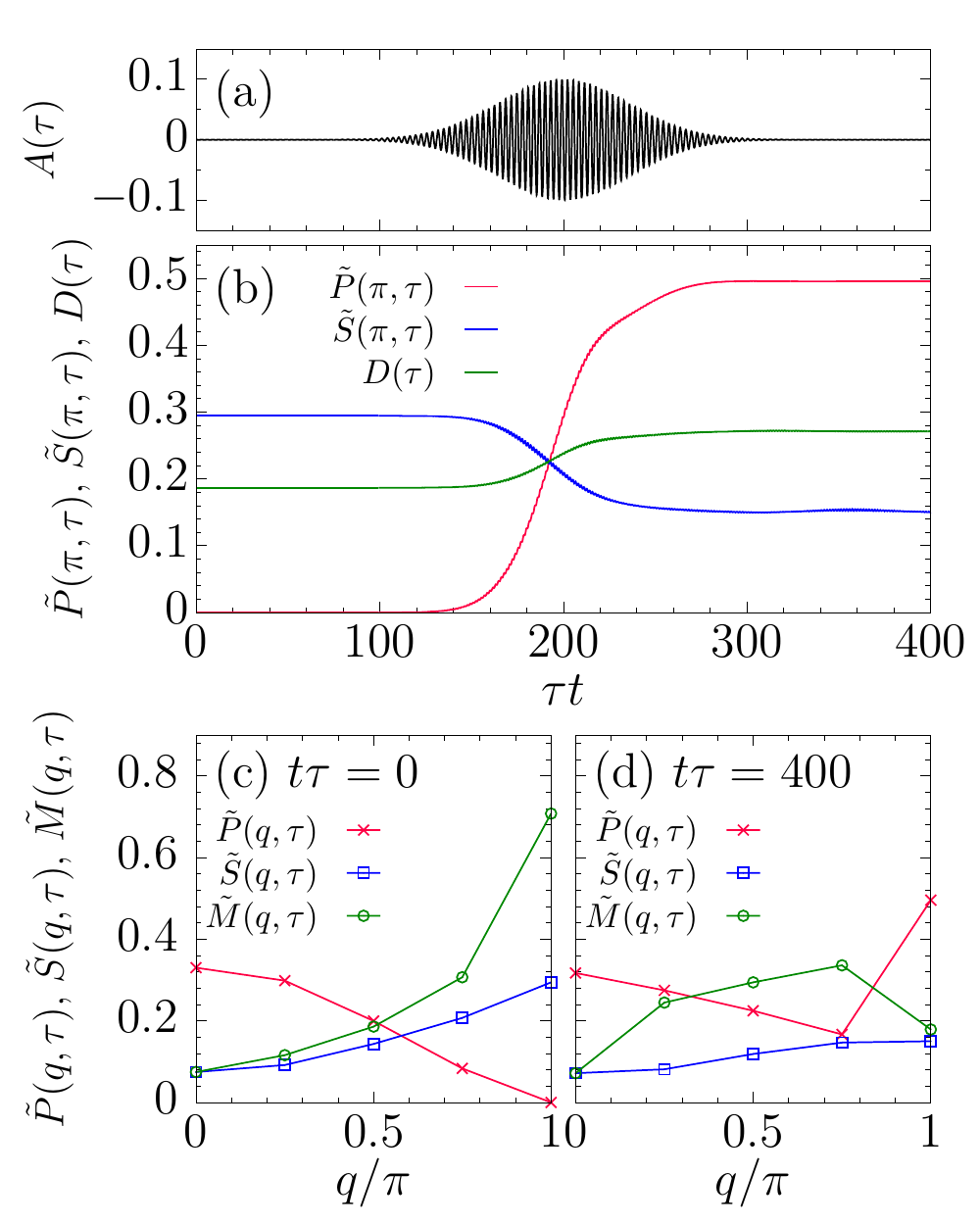}
  \caption{
  (a) Time-dependent external field $A(\tau)$ used here with $A_0 = 0.1$, $\omega_p = 2.05 t$,
    $\tau_c = 200/t$, and $\tau_w = 50/t$~\cite{note}. (b) Time evolution of 
    the on-site pair correlation function $\tilde{P}(q=\pi,\tau)$, the spin correlation function $\tilde{S}(q=\pi,\tau)$ for mobile electrons,
    and the double occupancy $D(\tau)$ for the 1D Kondo lattice model with $J=t$ and $L=8$ at half filling. 
    Momentum dependence of the on-site pair correlation function $\tilde{P}(q,\tau)$ and the spin 
    correlation functions $\tilde{S}(q,\tau)$ and $\tilde{M}(q,\tau)$ at (c) $\tau = 0$ and
    (d) $\tau = 400/t$ for the same model parameters used in (b).
    }
  \label{fig:tdep:cor}
\end{figure}

Figure~\ref{fig:tdep:cor} shows typical results of the time dependence of several correlation functions. 
These correlation functions include 
the on-site pair correlation function 
\begin{equation}
  \tilde{P}(q,\tau) = \frac{1}{L} \sum_{j=1}^L \sum_{j^{\prime}=1}^L
  {\rm e}^{-{\rm i} q (j-j^{\prime})}
  \langle \psi (\tau) \vert
  \hat{c}_{j\uparrow}^{\dagger} \hat{c}_{j\downarrow}^{\dagger}
  \hat{c}_{j^{\prime}\downarrow} \hat{c}_{j^{\prime}\uparrow}
  \vert \psi (\tau) \rangle, 
\end{equation}
the spin correlation function between mobile electrons 
\begin{equation}
  \tilde{S}(q,\tau) = \frac{1}{L} \sum_{j=1}^L \sum_{j^{\prime}=1}^L
  {\rm e}^{-{\rm i} q (j-j^{\prime})}
  \langle \psi (\tau) \vert
  \hat{S}^z_j 
  \hat{S}^z_{j^{\prime}}
  \vert \psi (\tau) \rangle, 
\end{equation}
the spin correlation function between localized spins
\begin{equation}
  \tilde{M}(q,\tau) = \frac{1}{L} \sum_{j=1}^L \sum_{j^{\prime}=1}^L
  {\rm e}^{-{\rm i} q (j-j^{\prime})}
  \langle \psi (\tau) \vert
  \hat{M}^z_j 
  \hat{M}^z_{j^{\prime}}
  \vert \psi (\tau) \rangle, 
\end{equation}
and the double occupancy  
\begin{equation}
  D (\tau)= \frac{1}{L} \sum_{j=1}^L 
  \langle \psi (\tau) \vert
  \hat{c}_{j\uparrow}^{\dagger} \hat{c}_{j\uparrow}
  \hat{c}_{j\downarrow}^{\dagger} \hat{c}_{j\downarrow}
  \vert \psi (\tau) \rangle.
\end{equation}
Notice first that since $\hat{\mathcal{H}}(\tau)$ is spin SU(2) symmetric even when $A(\tau)\ne0$, 
the spin correlation functions $\tilde{S}(q=0,\tau)$ and $\tilde{M}(q=0,\tau)$ defined above are exactly the same as those 
calculated for other spin components. Second, 
$\tilde{P}(q=\pi,\tau)$ corresponds to the correlation function for the $\eta$ pairing because 
$\tilde{P}(q=\pi,\tau)=\frac{1}{L}\langle \psi (\tau) \vert \hat{\eta}^+ \hat{\eta}^-\vert \psi (\tau) \rangle
=\frac{1}{L}\langle \psi (\tau) \vert  ( \hat{\mbox{\boldmath{$\eta$}}}^2 - \hat{\eta}_z^2 + \hat{\eta}_z )   \vert \psi (\tau) \rangle$. 
Third, $\tilde{S}(q=\pi,\tau)$ and $\tilde{M}(q=\pi,\tau)$ are the correlation functions for the antiferromagnetic ordering.

As shown in Figs.~\ref{fig:tdep:cor}(b) and \ref{fig:tdep:cor}(c), the $\eta$-pairing correlation $\tilde{P}(q=\pi,\tau=0)$ 
is exactly zero in the initial state. 
Since $\eta_z=0$ at half filling, this implies that the initial state has $\eta=0$. 
In contrast, the antiferromagnetic correlation is dominant 
in the initial state as we can observe in the correlation functions $\tilde{S}(q=\pi,\tau=0)$ and $\tilde{M}(q=\pi,\tau=0)$. 
We can also notice in Fig.~\ref{fig:tdep:cor}(b) that the double occupancy $D(\tau=0)$ in the initial state is around 0.18, 
much less than 0.25 expected for free electrons. The double occupancy is highly suppressed in the 
initial state because of the strong tendency toward the formation of local singlets.   
As the pulse is irradiated [also see Fig.~\ref{fig:tdep:cor}(a)], the pair correlation function $\tilde{P}(q=\pi,\tau)$ 
[the spin correlation function $\tilde{S}(q=\pi,\tau)$] gradually increases (decreases), 
and by the time the pulse irradiation is terminated, the pair correlation function $\tilde{P}(q=\pi,\tau)$ becomes 
dominant [see Figs.~\ref{fig:tdep:cor}(b) and ~\ref{fig:tdep:cor}(d)].
Note that the crossing of $\tilde{P}(q=\pi,\tau)$, $\tilde{S}(q=\pi,\tau)$ and $D(\tau)$
at $\tau  = 200/t$ in Fig.~\ref{fig:tdep:cor}(b) is simply accidental for this set of parameters.

\begin{figure}[htbp]
  \includegraphics[width=\hsize]{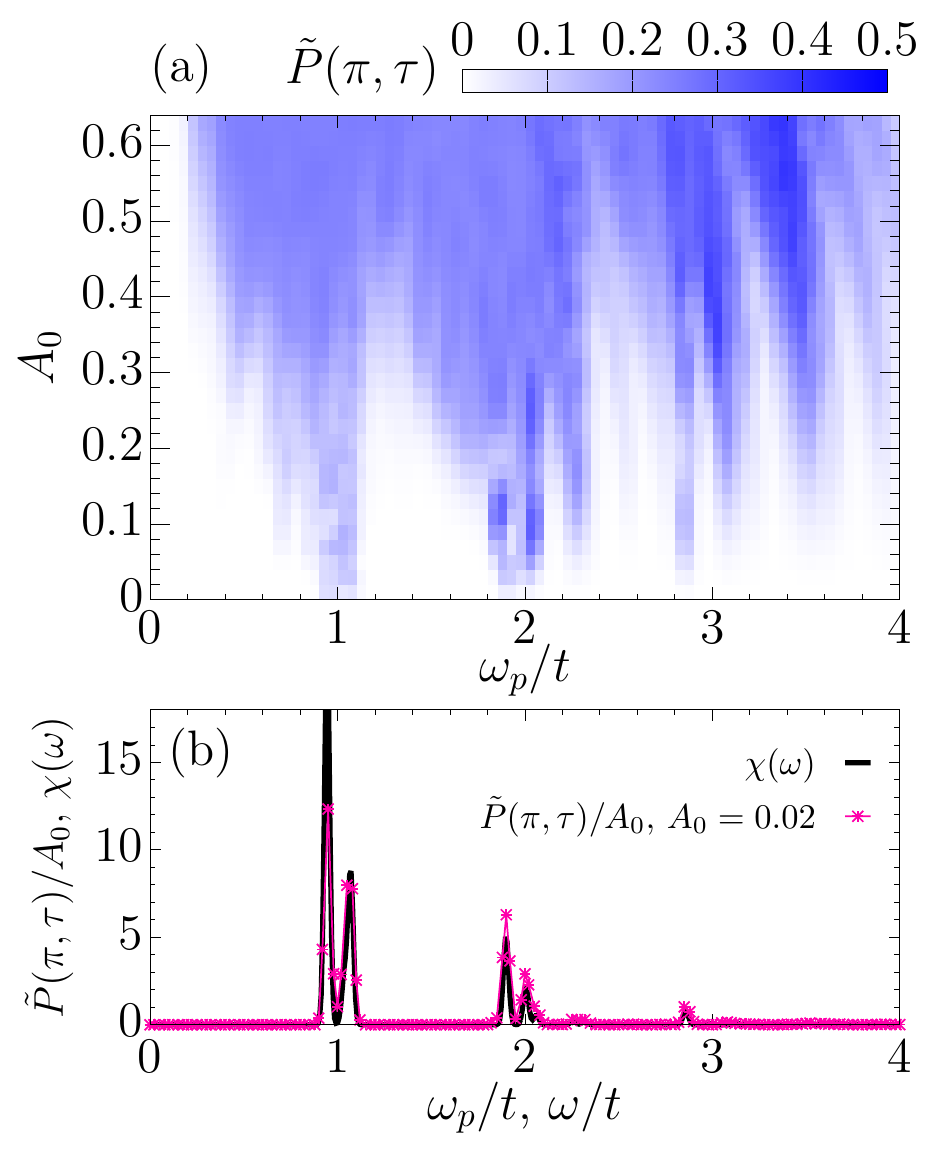}
  \caption{(a) Contour plot of the on-site pair correlation function $\tilde{P}(q=\pi,\tau)$
    at $\tau = 400/t$ with varying $\omega_p$ and $A_0$. 
    (b) On-site pair correlation function $\tilde{P}(\pi,\tau=400/t)$ as a function of $\omega_p$ for $A_0=0.02$ 
    and dynamical current correlation function $\chi(\omega)$ for the initial ground state. 
    $\varepsilon$ in $\chi(\omega)$ is $0.01t$. 
    The results are obtained for the 1D Kondo lattice model with $J=t$ and $L=8$ at half filling. 
    We use the external field $A(\tau)$ with $\tau_c=200/t$ and $\tau_\omega=50/t$. 
  }
  \label{fig:pmap}
\end{figure}

In order to find the optimal parameter set for the enhancement
of the $\eta$-pairing correlation, we show in Fig.~\ref{fig:pmap}(a) the contour plot of
$\tilde{P}(q=\pi,\tau)$ after the pulse irradiation at $\tau=400/t$
with different values of $A_0$ and $\omega_p$. 
As shown in Fig.~\ref{fig:pmap}(b), for small $A_0$, 
we find that $\tilde{P}(\pi,\tau=400/t)$ as a function of the frequencies $\omega_p$
almost coincides with the dynamical current correlation function
$\chi(\omega)$ for the initial ground state $ \vert \psi (0) \rangle=|\psi_0\rangle$ defined as
\begin{equation}
  \chi (\omega) = \langle \psi(0) \vert \hat{\mathcal{J}} \delta_{\varepsilon} (\omega - \hat{\mathcal{H}} + E_0 )
  \hat{\mathcal{J}} \vert \psi (0) \rangle,
  \label{eq:chi}
\end{equation}
where 
$\hat{\mathcal{J}}$ is the current operator given as
\begin{equation}
  \hat{\mathcal{J}} = - {\rm i}t \sum_{j=1}^L \sum_{\sigma = \uparrow,\downarrow} ( \hat{c}_{j\sigma}^{\dagger} \hat{c}_{j+1\sigma} - \hat{c}_{j+1\sigma}^{\dagger} \hat{c}_{j\sigma} )
\end{equation}
and 
\begin{equation}
  \delta_{\varepsilon}(\hat{\mathcal{X}}) = \frac{1}{\sqrt{2\pi \varepsilon^2}}
  \exp \left[ - \hat{\mathcal{X}}^2 / 2 \varepsilon^2 \right]
\end{equation}
for operator $\hat{\mathcal{X}}$, indicating that $\delta_{\varepsilon}(\hat{\mathcal{X}})$ approaches to the delta function 
in the limit of $\varepsilon \to 0^+$. 
We calculate $\chi(\omega)$ by using the method described in 
Appendix~\ref{appendix:spec}. 
The coincidence of these two quantities is expected 
from the facts that $\hat{\mathcal{J}} \vert \psi (0) \rangle$
has to be a state with $\eta = 1$, because $\hat{\mathcal{J}}$ is a rank-1 tensor operator, 
and the enhancement of $\tilde{P}(q=\pi,\tau=400/t)$ 
for small $A_0$ is essentially determined by the first order perturbation theory 
with $\cos A(\tau)\approx1$. 
Indeed, the $\omega_p$ dependence of $\tilde{P}(q=\pi,\tau=400/t)$ for large $A_0$
no longer follows $\chi(\omega)$ and the broad enhancement of the $\eta$-pairing correlation is found 
in a range of $0.3t\alt\omega_p\alt4t$.

\subsection{Distribution of $\mbox{\boldmath{$\eta$}}$-pairing eigenstates}

To investigate the distribution of $\eta$-pairing eigenstates in the photoexcited state $\vert \psi (\tau) \rangle$,
let us calculate the spectral function $P(\eta,\omega,\tau)$ given by
\begin{equation}
  P(\eta,\omega,\tau) = \langle \psi (\tau) \vert
  \hat{\mathcal{E}}_{\delta}(\eta) \delta_{\varepsilon}(\omega - \hat{\mathcal{H}} + E_0)
  \hat{\mathcal{E}}_{\delta}(\eta) \vert \psi (\tau) \rangle
  \label{eq:spec}
\end{equation}
where $\hat{\mathcal{E}}_{\delta}(\eta)$ is the projection operator
onto the subspace with a given value of $\eta$, i.e., 
\begin{equation}
  \hat{\mathcal{E}}_{\delta}(\eta) =
  \exp \left[
    - \left(
     \hat{\mbox{\boldmath{$\eta$}}}^2
        - \eta ( \eta + 1 ) \right)^2
    / \delta^2
    \right] 
    \label{eq:edelta}
\end{equation}
in the limit of $\delta\to0$. For a practical value of $\delta$ in the numerical calculations, we set 
$\delta$ as small as $1/\sqrt{5}$. 
The detail of the numerical implementation is described in Appendix~\ref{appendix:spec}.

Figure~\ref{fig:spec} shows typical results of the spectral function $P(\eta,\omega,\tau)$ 
calculated at $\tau=400/t$. 
We find that finite intensities appear at $\omega \sim 2n \omega_p$ 
for $\eta$ even and $\omega \sim (2n+1) \omega_p$ for $\eta$ odd, where $n$ is non-negative integer. 
This is in good accordance with the result for the time-dependent perturbation theory described in Sec.~\ref{sec:perturb}.
Notice that the finite contribution with $\eta > 1$ cannot be explained 
by the first-order perturbation process since the rank-1 tensor operator can change
the value of $\eta$ by 1, indicating the importance of the non-linear processes. 
These eigenstates with $\eta$ finite are responsible for the enhancement of the $\eta$-pairing correlation in the photoinduced state.

\begin{figure}[htbp]
  \includegraphics[width=\hsize]{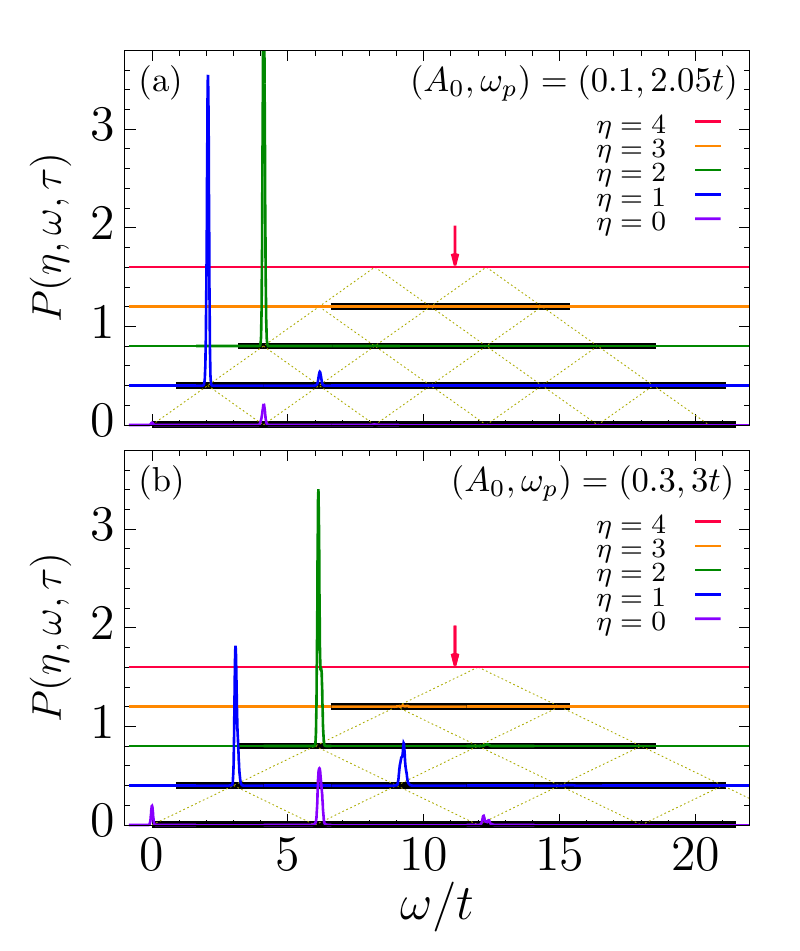}
  \caption{Spectral function $P(\eta,\omega,\tau)$ in the photoexcited state $\vert \psi (\tau) \rangle$ at $\tau=400/t$ 
    for the half-filled 1D Kondo lattice model with $J=t$ and $L=8$ under  
    the external field $A(\tau)$ with (a) $A_0 = 0.1$ and $\omega_p = 2.05t$, and (b) $A_0 = 0.3$ and $\omega_p = 3t$. 
    The other parameters for $A(\tau)$ are $\tau_c=200/t$ and $\tau_\omega=50/t$. 
    For visibility, the spectral functions with different values of $\eta$ are shifted vertically.  
    Black solid lines indicate the energy region
    where the eigenstates of $\cal\hat{H}$ exist for each $\eta$. 
    Red arrows indicate the excitation energy of the vacuum state $\vert {\rm vac} \rangle$. 
    Crossing points between dashed lines and black solid lines indicate 
    $\omega = 2 n \omega_p$ for $\eta$ even and $\omega = (2n+1) \omega_p$ for $\eta$ odd, 
    where $n = 0,1,2,\cdots$. 
  }
  \label{fig:spec}
\end{figure}

\subsection{Dynamical spin correlation}

\begin{figure}[htbp]
  \includegraphics[width=\hsize]{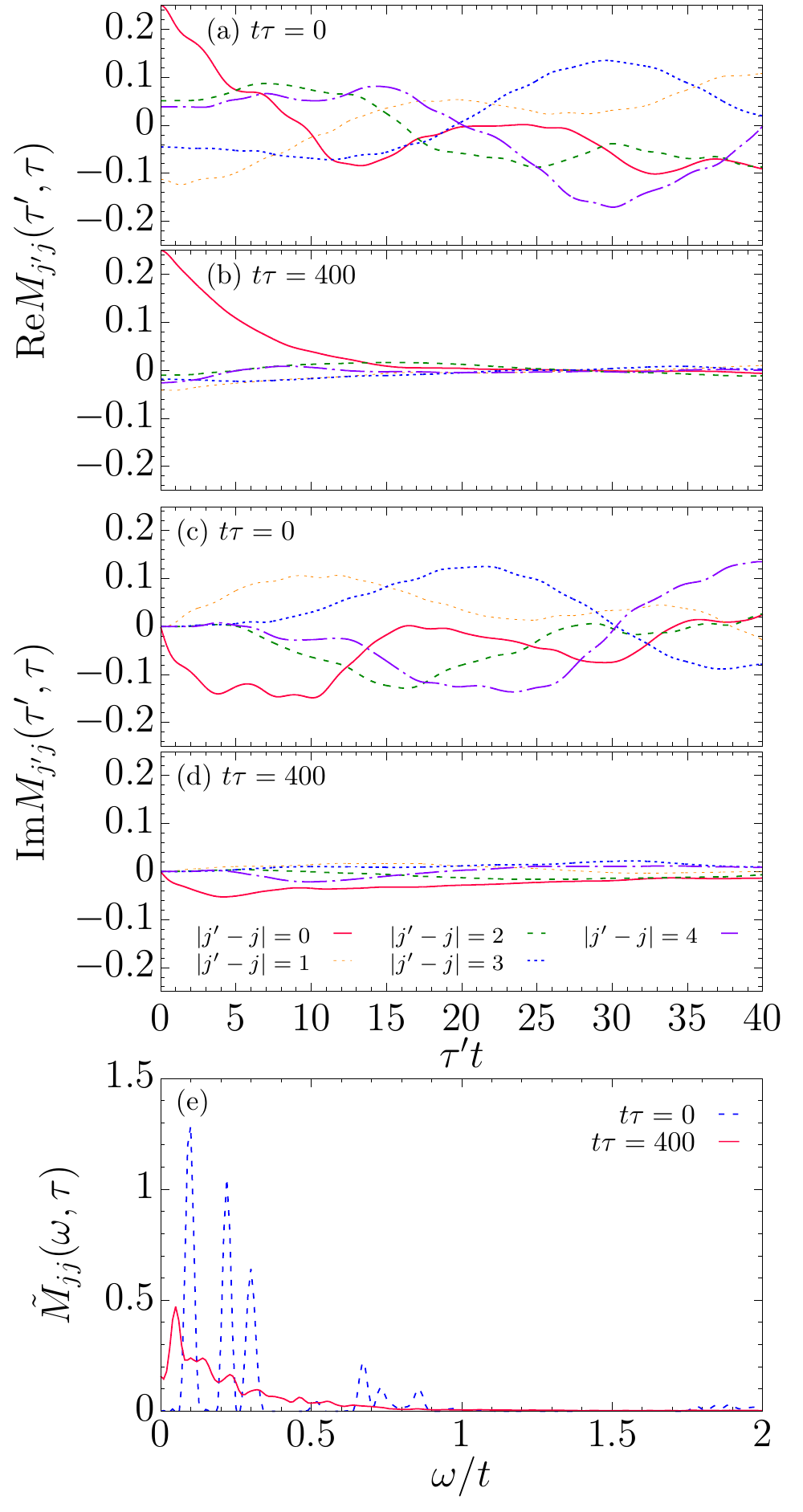}
  \caption{
  (a) (b) Real part of dynamical correlation functions $M_{j^{\prime}j}(\tau^{\prime},\tau)$ for (a) the initial state at $\tau = 0$
  and (b) the time-evolved state after the pulse irradiation at $\tau = 400/t$.
  (c) (d) Imaginary part of $M_{j^{\prime}j}(\tau^{\prime},\tau)$ for (c) the initial state $\tau = 0$
  and (d) the time-evolved state after the pulse irradiation at $\tau = 400/t$.
  (e) Frequency dependent dynamical correlation function
  $\tilde{M}_{jj}(\omega,\tau)$ at $\tau=0$ and $400/t$. 
  The results are obtained for the half-filled 1D Kondo lattice model with $J=t$ and $L=8$ under the external field 
  $A(\tau)$ with $A_0 = 0.38$, $\omega_p  = 3/t$, $\tau_w = 50/t$, and $\tau_c = 200/t$. 
  }
  \label{fig:locm}
\end{figure}

In order to examine the effects of the photoinduced $\eta$ pairs 
on the localized spins, here we calculate the dynamical correlation function for the localized spins defined as 
\begin{equation}
  M_{j^{\prime}j} (\tau^{\prime},\tau) =
  \langle \psi(\tau) \vert \hat{M}^z_{j^{\prime}}(\tau^{\prime}) \hat{M}^z_{j} \vert \psi (\tau) \rangle,
\end{equation}
where 
\begin{equation}
  \hat{M}^z_{j}(\tau) = {\rm e}^{{\rm i}\hat{\mathcal{H}} \tau} \hat{M}^z_{j} {\rm e}^{-{\rm i}\hat{\mathcal{H}}\tau}. 
\end{equation}
Here, $M_{j^{\prime}j}(\tau^{\prime},\tau)$ is a quantity indicating how the spin $\hat{M}^z_{j}$ for the state $\vert \psi(\tau) \rangle$
is correlated to the spin $\hat{M}_{j^{\prime}}^z$ after the time $\tau^{\prime}$.
Figures~\ref{fig:locm}(a) and (c) show the results of $M^z_{j^{\prime}j}(\tau^{\prime},\tau)$ for the initial state at $\tau = 0$,
which is the ground state of the half-filled 1D Kondo lattice model $\hat{\mathcal{H}}$. 
Recalling that the ground state is an insulating state with a finite spin gap
due to the formation of local singlets, the low-lying spin excitations are 
described by the triplon-like excitations similar to the excitations in the valence bond solids~\cite{PhysRevB.41.9323,PhysRevB.49.8901,PhysRevB.64.092406}.
In the Kondo lattice model, the triplon is a local object
composed of a conduction electron and a localized spin, and 
this picture is more preferable for large $J/t$ because
for a small or moderate value of $J/t$ the triplon is likely a spatially more extended object.
Therefore, in a small $J/t$ region, we expect the spin excitations similar to those found in the case where 
the antiferromagnetic correlation is dominant. 
However, these different behaviors are not distinguishable in our simulation
using the limited size of clusters. Instead, we only find in Fig.~\ref{fig:locm}(a) that
the correlation is antiferromagnetic at $\tau^{\prime}=0$ and
starts to oscillate gradually from the nearest to distant sites,
which is characteristics for the ballistic dynamics 
in systems with strong antiferromagnetic correlation.

In contrast, we find in Figs.~\ref{fig:locm}(b) and \ref{fig:locm}(d) that $M_{j^{\prime}j}(\tau^{\prime},\tau)$
after the pulse irradiation behaves more diffusive. The dynamical correlations
not only for the same site $j=j^{\prime}$ but also for the distant sites decrease in time,
and the oscillatory behavior is no longer remarkable.
Such a diffusive nature indicates
the softening of the spectral function because the diffusive nature suggests
a quadratic form of the energy dispersion with respect to momentum.
In Fig.~\ref{fig:locm}(c), 
we display the frequency dependent dynamical correlation function $\tilde{M}_{jj}(\omega,\tau)$ obtained by  
the Fourier transform of $M_{jj}(\tau^{\prime},\tau)$, i.e., 
\begin{equation}
  \tilde{M}_{jj}(\omega,\tau) = \frac{1}{T} \int_{0}^{T} {\rm d}\tau^{\prime} g(\tau')
  M_{jj}(\tau^{\prime},\tau) {\rm e}^{{\rm i}\omega \tau^{\prime}},
\end{equation}
where we set $T = 200/t$ and multiply the integrand by 
a contour $g(\tau)=\left(1 + \cos(\pi \tau/T)\right)/2$
which makes delta peaks broad, for obtaining a smooth function~\cite{arxiv:1901.07751}. 
We find that the the main peaks in $\tilde{M}_{jj}(\omega,\tau)$ after the pulse irradiation at $\tau = 400/t$ 
shift to lower energies as compared to the spectrum before the pulse irradiation at $\tau = 0$.

The diffusive behavior of the dynamical spin correlation function is 
understood as a consequence of the generation of $\eta$ pairs, which 
leads to the decoupling between the localized spins and the mobile electrons, 
as discussed for Yang's state in Sec.~\ref{sec:etaop}. 
Here, we show that this picture is related to the photodoping mechanism~\cite{Maeshima2005,Werner2018,Werner2019,Werner2019B,Li2019}.
To this end, we introduce $\vert \phi_{\alpha,\eta,\eta_z} \rangle$ being
a state for $(\eta,\eta_z)$-sector 
to decompose the photoexcited state $\vert \psi (\tau) \rangle$ as 
\begin{equation}
  \vert \psi (\tau) \rangle = \sum_{\alpha,\eta} c_{\alpha\eta}(\tau) \vert \phi_{\alpha,\eta,\eta_z} \rangle,
\end{equation}
where index $\alpha$ is introduced to distinguish states with the same value of $\eta$. Note that $\eta_z$ is fixed to be zero at half filling. 
Let $M^{(\alpha,\eta,\eta_z),(\alpha',\eta^{\prime},\eta_z^{\prime})}_{jj^{\prime}}(\tau^{\prime})$ be
the dynamical correlation function defined by
\begin{equation}
  M^{(\alpha,\eta,\eta_z),(\alpha',\eta^{\prime},\eta_z^{\prime})}_{j^{\prime}j}(\tau^{\prime}) = 
  \langle \phi_{\alpha,\eta,\eta_z} \vert \hat{M}^z_{j^{\prime}}(\tau^{\prime}) \hat{M}^z_j
  \vert \phi_{\alpha',\eta^{\prime},\eta_z^{\prime}}  \rangle.
\end{equation}
Notice that $\vert \phi_{\alpha,\eta,\eta_z} \rangle$ is not necessarily an energy eigenstate of $\hat{\mathcal{H}}$
but an eigenstate of $\hat{\mbox{\boldmath{$\eta$}}}^2$ and $\hat{\eta}_z$. 
Since $\hat{\eta}^+$ and $\hat{\eta}^-$ commute with $\hat{M}^z_i$ as well as $\hat{\mathcal{H}}$, 
$ \hat{M}^z_{j^{\prime}}(\tau^{\prime}) \hat{M}^z_j$ is considered as a rank-0 tensor operator
for the pseudo-spin operators $\hat{\eta}_{\mu}$ ($\mu =x,y,z$).
Therefore, applying the Wigner-Eckert theorem in Eq.~(\ref{eq:wet}), we obtain
\begin{equation}
  M^{(\alpha,\eta,\eta_z),(\alpha',\eta^{\prime},\eta_z^{\prime})}_{j^{\prime}j}(\tau^{\prime}) =
  \delta_{\eta,\eta^{\prime}}
  \delta_{\eta_z,\eta_z^{\prime}}
  M^{(\alpha,\eta,-\eta),(\alpha',\eta,-\eta)}_{j^{\prime}j} (\tau^{\prime}) 
  \label{eq:locm:rule:1}
\end{equation}
because
\begin{equation}
  \langle \eta,\eta_z 0 0 \vert \eta^{\prime} \eta_z^{\prime} \rangle =
  \delta_{\eta,\eta^{\prime}}
  \delta_{\eta_z,\eta_z^{\prime}}. 
\end{equation}
Equation~(\ref{eq:locm:rule:1}) suggests
that the dynamical spin correlation function $M^{(\alpha,\eta,\eta_z),(\alpha',\eta^{\prime},\eta_z)}_{j^{\prime}j}(\tau^{\prime})$
for electron number $N = L + 2 \eta_z$ exactly coincides to that for $N = L - 2 \eta$. 
Note that $\vert \phi_{\alpha,\eta,\eta_z=-\eta} \rangle$ is a lowest weight state~\cite{EFGetal05} because 
$\hat{\eta}^- \vert \phi_{\alpha,\eta,-\eta}\rangle=0$, implying that there are no $\eta$ pairs in $\vert \phi_{\alpha,\eta,-\eta} \rangle$.  
Moreover, by using Eq.~(\ref{eq:locm:rule:1}),
the dynamical spin correlation function after the pulse irradiation 
can be represented as
\begin{equation}
  M_{j^{\prime}j}(\tau^{\prime},\tau) = \sum_{\alpha,\alpha',\eta} c_{\alpha\eta}^*(\tau) c_{\alpha'\eta}(\tau) 
  M^{(\alpha,\eta,-\eta),(\alpha',\eta,-\eta)}_{j^{\prime}j} (\tau^{\prime}),
  \label{eq:locm:rule:2}
\end{equation}
suggesting that the spin dynamical correlation function at half filling (i.e., $\eta_z=0$) can be
represented as a simple sum of the spin dynamical correlation functions for the hole-doped systems.

We thus find that the photogeneration of $\eta$ pairs 
is inseparably related to photodoping when we consider the spin dynamics. 
This is expected because 
the energy eigenstate with quantum number $\eta=-\eta_z=\frac{1}{2}(N-2N_\eta-L)$ span
the subspace of the $N-2N_\eta$ electron system and
the multiplication of $\hat{\eta}^+$ operator by $N_\eta$ times to this state 
replaces $N_\eta$ empty sites with doubly occupied sites, 
which is an energy eigenstate with $N$ electrons, having the same energy eigenvalue, 
and does not affect the properties of spin degrees of freedom. 
Considering that the ground state of the 1D Kondo lattice model varies from a paramagnetic phase to 
a ferromagnetic phase with increasing the hole concentration~\cite{PhysRevB.47.8345}, 
we can understand that the photogeneration of $\eta$ pairs, which is essentially the  
in-situ doping, changes drastically the spin dynamics of the initial ground state at half filling.

Finally, we also notice in Fig.~\ref{fig:locm}(b) that 
the equal-time correlation $M_{j^{\prime}j}(\tau'=0,\tau)$ is less dependent on
the distance $\vert j^{\prime}-j \vert\,(\ne0)$. This tendency 
is explained by the dephasing mechanism recently proposed in Ref.~\cite{PhysRevLett.123.030603}. 
This mechanism states that in a system with SU(2)$\times$SU(2) symmetry, e.g, having both spin- and $\eta$-SU(2) symmetries, 
if we apply a Floquet-type time-dependent perturbation that commutes with one of the SU(2) symmetries but breaks the other symmetry, 
the correlation function composed of the local operators represented by generators of
the SU(2) symmetry that commutes with the perturbation becomes spatially uniform in a steady state.
For example, in Ref.~\cite{PhysRevLett.123.030603}, they have demonstrated
in the Hubbard model that by applying the time-dependent perturbation $h(\tau) \sum_{j} \hat{S}_j^z$, 
the correlation function 
$\langle \hat{c}_{j\downarrow} \hat{c}_{j\uparrow} \hat{c}_{j^{\prime}\uparrow}^{\dagger} \hat{c}_{j^{\prime}\downarrow}^{\dagger} \rangle$
in a steady state becomes spatially uniform. Notice that $h(\tau)\sum_j \hat{S}_j^z$ breaks the spin-SU(2) symmetry and 
changes the quantum number of the total spin, 
but commutes with $\hat{c}_{j\downarrow} \hat{c}_{j\uparrow}$. 
This dephasing mechanism can be applied to our case. 
The time-dependent perturbation is $\hat{\mathcal{V}}(\tau)$ that breaks the $\eta$-SU(2) symmetry and changes the quantum number 
of $ \hat{\mbox{\boldmath{$\eta$}}}^2 $, but commutes with $\hat{M}_j^z$. Therefore, 
the correlation function $M_{jj^{\prime}}(\tau'=0,\tau)$ becomes spatially uniform in a steady state, 
as found in Fig.~\ref{fig:locm}(b).  
We should note that the discussion given above is not limited for 
the localized spins but also applied to the electron spins in the conduction band.

\section{\label{sec:summary}Summary and Discussion}

By using the time-dependent exact diagonalization technique, 
we have shown that the pulse irradiation can generate $\eta$ pairs and thus induce the enhancement of the pair-density-wave-like 
superconducting correlation 
in the ground state of the 1D Kondo lattice model at half filling. 
The $\eta$-pairing states are preferentially generated by the optical field because of the symmetry associated with the 
$\eta$-pairing operators $\hat{\mbox{\boldmath{$\eta$}}} = (\hat{\eta}_x, \hat{\eta}_y, \hat{\eta}_z)$ that satisfy the 
SU(2) commutation relations. This was also analytically shown using the time-dependent perturbation theory, by which 
the selection rule becomes apparent. 
We have furthermore investigated the effect on the localized spin degrees of
freedom in the photoexcited state and  
found that the spin dynamics becomes diffusive after the pulse irradiation. 
This is understood because the generation of $\eta$ pairs decouple locally the exchange interaction, 
which is essentially equivalent to the effective hole doping.

The numerical simulation in Sec.~\ref{sec:result} is 
for the 1D Kondo lattice model with the antiferromagnetic exchange interaction. 
However, the symmetry analysis given in Sec.~\ref{sec:method} is applicable for any spatial dimension 
as long as the system is bipartite. It is also obvious that the sign of
the exchange interaction does not affected the discussion in Sec.~\ref{sec:method}. 
Moreover, since the pseudo-spin operators $\hat{\mbox{\boldmath{$\eta$}}} = (\hat{\eta}_x, \hat{\eta}_y, \hat{\eta}_z)$ 
are defined only for the mobile electrons, the symmetry analysis given in Sec.~\ref{sec:method} is still correct 
even when we add any exchange interaction term between the localized spins 
\begin{equation}
  \sum_{j,j^{\prime}} \bar{J}_{jj^{\prime}} \hat{{\bm M}}_j \cdot \hat{{\bm M}}_{j^{\prime}}
\end{equation}
which are not necessarily in the bipartite structure. Therefore, 
we can prepare various initial states with different spin structures, including the Haldane phase in the 1D 
system with $J < 0$ as well as a quantum spin liquid for the frustrated exchange coupling $\bar{J}_{jj^{\prime}}$.
We have implicitly assumed that the localized spins $\hat{{\bm M}}_j$ are spin 1/2. 
However, this assumption is not necessary for the symmetry 
analysis in Sec.~\ref{sec:method}. The $\eta$-pairing operators 
$\hat{\mbox{\boldmath{$\eta$}}} = (\hat{\eta}_x, \hat{\eta}_y, \hat{\eta}_z)$ and the $\eta$-pairing states 
are still well defined even when we consider the Kondo lattice model with the classical localized spins.

  Another model related to this study is the periodic Anderson model described by the following Hamiltonian: 
  \begin{equation}
    \hat{\mathcal{H}}_{\rm PA} = \hat{\mathcal{H}}_t + \hat{\mathcal{H}}_V + \hat{\mathcal{H}}_U
    \label{eq:model:anderson}
  \end{equation}
  where
  \begin{equation}
    \hat{\mathcal{H}}_V = V \sum_{j} \sum_{\sigma = \uparrow,\downarrow} ( \hat{c}_{j\sigma}^{\dagger} \hat{d}_{j\sigma} + \hat{d}_{j\sigma}^{\dagger} \hat{c}_{\sigma}) 
  \end{equation}
  and
  \begin{equation}
    \hat{\mathcal{H}}_U = U \sum_{j}
    ( \hat{d}_{j\uparrow}^{\dagger} \hat{d}_{j\uparrow} - \frac{1}{2} )
    ( \hat{d}_{j\downarrow}^{\dagger} \hat{d}_{j\downarrow} - \frac{1}{2} ) .
  \end{equation}
  Here, $\hat{\mathcal{H}}_t$ is defined in Eq.~(\ref{eq:ham:hop}) and 
  $\hat{d}_{j\sigma}$ ($\hat{d}_{j\sigma}^{\dagger}$) denotes the annihilation (creation)
  operator of a localized electron with spin $\sigma$ ($=\uparrow,\downarrow$) at site $j$.
  It is well known that the Kondo lattice model is the effective low-energy model of the periodic Anderson model in the limit of 
  $U \to \infty$.
  We should note that, in this case, the $\eta$-pairing operators $\hat{\mbox{\boldmath{$\eta$}}} = (\hat{\eta}_x, \hat{\eta}_y, \hat{\eta}_z)$ 
  defined in Eqs.~(\ref{eq:eta_x}), (\ref{eq:eta_y}), and (\ref{eq:eta_z}) do not commute with $\hat{\mathcal{H}}_{\rm PA}$, but  
  their definition has to be extended as follows:
  \begin{eqnarray}
    \hat{\eta}_x^{({\rm PA})} &= & \hat{\eta}_x
     - \frac{1}{2} \sum_{j} {\rm e}^{{\rm i}\phi_j} ( \hat{d}_{j\uparrow}^{\dagger} \hat{d}_{j\downarrow}^{\dagger} + \hat{d}_{j\downarrow} \hat{d}_{j\uparrow} ), \label{eq:ex:eta:x} \\
    \hat{\eta}_y^{({\rm PA})} &= & \hat{\eta}_y
     - \frac{1}{2{\rm i}} \sum_{j} {\rm e}^{{\rm i}\phi_j} ( \hat{d}_{j\uparrow}^{\dagger} \hat{d}_{j\downarrow}^{\dagger} - \hat{d}_{j\downarrow} \hat{d}_{j\uparrow} ), \label{eq:ex:eta:y} \\
    \hat{\eta}_z^{({\rm PA})}  &= & \hat{\eta}_z
     + \frac{1}{2} \sum_{j} ( \hat{d}_{j\uparrow}^{\dagger} \hat{d}_{j\uparrow} + \hat{d}_{j\downarrow}^{\dagger} \hat{d}_{j\downarrow} - 1 ).  
     \label{eq:ex:eta:z}
  \end{eqnarray}
  Notice that these operators satisfy the SU(2) commutation relations and commute with $\hat{\mathcal{H}}_{\rm PA}$ 
  for any $U$, including the case when $U$ is negative. 
  
  The negative-$U$ periodic Anderson model 
  has been considered to discuss the charge Kondo effect
  for materials containing valence skipping elements~\cite{matsuura2012}.
  The strong coupling limit of the negative-$U$ periodic Anderson model is described by the charge Kondo lattice model:
  \begin{equation}
    \hat{\mathcal{H}}_{c{\rm KL}} = \hat{\mathcal{H}}_t + \hat{\mathcal{H}}_{\tilde{J}}
  \end{equation}
  where
  \begin{equation}
    \hat{\mathcal{H}}_{\tilde{J}} = \tilde{J} \sum_{j} \hat{\mbox{\boldmath{$\eta$}}}_{j} \cdot
    \hat{\mbox{\boldmath{$\eta$}}}_{jd}.
  \end{equation}
  Here, $\hat{\mbox{\boldmath{$\eta$}}}_{j} = (\hat{\eta}_{j}^x, \hat{\eta}_{j}^y, \hat{\eta}_{j}^z)$ and 
  $\hat{\mbox{\boldmath{$\eta$}}}_{jd} = (\hat{\eta}_{jd}^x, \hat{\eta}_{jd}^y, \hat{\eta}_{jd}^z)$ 
  represent the local $\eta$-pairing operators for 
  the conduction and localized electrons, respectively, 
  given by
  \begin{align}
    \hat{\eta}^x_{j} = & \frac{1}{2} {\rm e}^{{\rm i}\phi_j} (\hat{c}_{j\uparrow}^{\dagger}\hat{c}_{j\downarrow}^{\dagger} + \hat{c}_{j\downarrow}\hat{c}_{j\uparrow}), \\
    \hat{\eta}^y_{j} = & \frac{1}{2{\rm i}} {\rm e}^{{\rm i}\phi_j} (\hat{c}_{j\uparrow}^{\dagger}\hat{c}_{j\downarrow}^{\dagger} - \hat{c}_{j\downarrow}\hat{c}_{j\uparrow}), \\
    \hat{\eta}^z_{j} = & \frac{1}{2}  (\hat{c}_{j\uparrow}^{\dagger}\hat{c}_{j\uparrow} + \hat{c}_{j\downarrow}^{\dagger} \hat{c}_{j\downarrow} - 1),
  \end{align}
  and 
  \begin{align}
    \hat{\eta}^x_{jd} = & - \frac{1}{2} {\rm e}^{{\rm i}\phi_j} ( \hat{d}_{j\uparrow}^{\dagger} \hat{d}_{j\downarrow}^{\dagger} + \hat{d}_{j\downarrow} \hat{d}_{j\uparrow} ), \\
    \hat{\eta}^y_{jd} = & - \frac{1}{2{\rm i}} {\rm e}^{{\rm i}\phi_j} ( \hat{d}_{j\uparrow}^{\dagger} \hat{d}_{j\downarrow}^{\dagger} - \hat{d}_{j\downarrow} \hat{d}_{j\uparrow} ), \\
    \hat{\eta}^z_{jd} = &  \frac{1}{2} ( \hat{d}_{j\uparrow}^{\dagger} \hat{d}_{j\uparrow} + \hat{d}_{j\downarrow}^{\dagger} \hat{d}_{j\downarrow} - 1 ),
  \end{align}
  and $\tilde{J} = 8 V^2/|U|$. 
  Notice that these local $\eta$-pairing operators also satisfy the SU(2) commutation relations among themselves, 
  i.e., $[ \hat{\eta}_j^{\mu}, \hat{\eta}_{j^{\prime}}^{\nu} ] =
       {\rm i}\delta_{jj^{\prime}}\sum_\lambda \varepsilon_{\mu \nu \lambda} \hat{\eta}_j^{\lambda}$ 
   and $[ \hat{\eta}_{jd}^{\mu}, \hat{\eta}_{j^{\prime}d}^{\nu} ] =
       {\rm i}\delta_{jj^{\prime}}\sum_\lambda \varepsilon_{\mu \nu \lambda} \hat{\eta}_{jd}^{\lambda}$. 
  In this case, even although the total charge of the conduction electrons fluctuates, 
  the extended $\eta$-pairing operators in Eqs.~(\ref{eq:ex:eta:x}), (\ref{eq:ex:eta:y}), 
  and (\ref{eq:ex:eta:z}) still commute with the charge Kondo lattice Hamiltonian $\hat{\mathcal{H}}_{c{\rm KL}}$. 

The photoexcitation of these systems are highly interesting and
the research along this line is now in progress.

\section*{Acknowledgement}
The authors are grateful to T. Kaneko for fruitful discussion. 
The calculation has been performed on the RIKEN supercomputer system (HOKUSAI GreatWave). 
This work was supported in part by JST PRESTO (No.~JPMJPR191B), Japan, and also by 
Grant-in-Aid for Scientific Research (B) (No.~JP18H01183) from MEXT, Japan.

\appendix

\section{\label{appendix:spec}Technical details of numerical calculations}

This appendix summarizes the numerical methods to calculate 
the dynamical current correlation function $\chi(\omega)$ in Eq.~(\ref{eq:chi}) and 
the spectral function $P(\eta,\omega,\tau)$ in Eq.~(\ref{eq:spec}).

Let us first describe the method to calculate
the following projected state 
\begin{equation}
  \vert \psi_{\eta}(\tau) \rangle = \hat{\mathcal{E}}_{\delta}(\eta) \vert \psi (\tau) \rangle, 
\end{equation}
which appears in Eq.~(\ref{eq:spec}).
We first divide the exponential operator $ \hat{\mathcal{E}}_{\delta}(\eta)$ in Eq.~(\ref{eq:edelta}) into many slices:
\begin{equation}
  \hat{\mathcal{E}}_{\delta}(\eta) =
  \left[
    \hat{\mathcal{E}}_{\sqrt{M}\delta}(\eta)
    \right]^M. 
\end{equation}
We then multiply $\hat{\mathcal{E}}_{\sqrt{M}\delta}(\eta)$ sequentially to $\vert \psi (\tau) \rangle$ as 
\begin{equation}
  \vert v_m \rangle = \hat{\mathcal{E}}_{\sqrt{M}\delta}(\eta) \vert v_{m-1} \rangle
\end{equation}
for $m = 1,2,\cdots, M$ with
\begin{equation}
  \begin{split}
 & \vert v_0 \rangle = \vert \psi (\tau) \rangle, \\ 
 & \vert v_M \rangle = \vert \psi_{\eta}(\tau) \rangle. 
  \end{split}
\end{equation}
At each step of $m$, we use the Taylor expansion of the exponential operator $\hat{\mathcal{E}}_{\sqrt{M}\delta}(\eta)$, 
similar to the case for the time-evolution operator in Eqs.~(\ref{eq:time-evolution})--(\ref{eq:teite}).
We find that this is numerically stable for any case studied here.

Next, we should notice that the dynamical current correlation function $\chi(\omega)$ and 
the spectral function $P(\eta,\omega,\tau)$ have the following form:
\begin{equation}
  S(\omega) = \langle \psi \vert
  \delta_{\varepsilon} ( \hat{\mathcal{H}} - \omega + E_0 )
  \vert \psi \rangle
\end{equation}
where $\vert \psi \rangle = \hat{\mathcal{J}} \vert \psi (0) \rangle$ for $\chi (\omega)$ and 
$\vert \psi \rangle = \hat{\mathcal{E}}_{\delta}(\eta) \vert \psi (\tau) \rangle$ for $P(\eta,\omega,\tau)$.
To calculate $S(\omega)$,
we use the following formula: 
\begin{equation}
  S(\omega) = \frac{1}{\sqrt{2\pi \varepsilon^2}}
  \sum_{k=1}^K {\rm e}^{- (\xi_k - \omega + E_0)^2 / 2 \varepsilon^2}
  \vert \langle e_1 \vert \xi_k \rangle \vert^2. 
  \label{eq:s_w}
\end{equation}
Here, $\vert e_k \rangle$ with $k=1,2,\cdots,K$ is a set of 
the orthonormalized bases (i.e., $\langle e_k|e_{k'}\rangle=\delta_{k,k'}$) generated by the Lanczos procedure
\begin{equation}
  \begin{split}
    & \vert e_1 \rangle = \vert \psi \rangle, \\
    & \beta_1 \vert e_2 \rangle = \hat{\mathcal{H}} \vert e_1 \rangle - \alpha_1 \vert e_1 \rangle, \\
    & \beta_k \vert e_{k+1} \rangle = \hat{\mathcal{H}} \vert e_{k} \rangle - \alpha_k \vert e_{k} \rangle -
    \beta_{k-1} \vert e_{k-1} \rangle \,\,\,\,(\text{for}\,k\ge2), 
  \end{split}
  \label{eq:lanczos}
\end{equation}
where $\alpha_k = \langle e_k \vert \hat{\mathcal{H}} \vert e_k \rangle$.
Notice that the coefficients $\alpha_k$ and $\beta_k$ correspond
to the matrix elements of Hamiltonian $\hat{\mathcal{H}}$ taken 
in the reduced Hilbert space spanned by the basis set $\{\vert e_k \rangle\}$. 
$\xi_k$ and $\vert \xi_k \rangle$ with $k=1,2,\cdots,K$ in Eq.~(\ref{eq:s_w}) 
are the approximate eigenvalues and eigenstates of $\hat{\mathcal{H}}$, respectively, obtained
by diagonalizing the tridiagonal matrix constructed via the Lanczos iteration in Eq.~(\ref{eq:lanczos}). 
Note that $\langle e_1 \vert \xi_k \rangle$ is the first element of $\vert \xi_k \rangle$ represented in the basis set of 
$\{|e_k\rangle\}$. 
$\xi_k$ yields the peak positions 
in the spectral function $S(\omega)$, which coincide with the pole positions 
obtained by the continued fraction technique~\cite{RevModPhys.66.763}.

\bibliography{KL_eta}

\begin{thebibliography}{45}
\bibitem [{not()}]{note}%
  \BibitemOpen
  \href@noop {} {}\bibinfo {note} {The pulse described by this external field
  $A(\tau)$ has the pulse width (full width at half maximum) of approximately
  150~fs and introduces the time-dependent electric field $E(\tau)=-\partial
  A(\tau)/\partial \tau$ with the maximum electric field as large as 2~MV/cm,
  assuming that $t = 0.5$~eV and a lattice constant $a= 5$~\AA}\BibitemShut
  {NoStop}%
\end{thebibliography}

\end{document}